\documentclass[english]{iopart}

\begin{document}

\title[Series expansion method and ODE of $\chi^{(3)}$ ]
{\Large
Square lattice Ising model susceptibility: Series expansion
method and differential equation for $\chi^{(3)}$}
 
\author{ 
N. Zenine$^\S$, S. Boukraa$^\dag$, S. Hassani$^\S$ and
J.-M. Maillard$^\ddag$}
\address{\S  C.R.N.A.,
Bld Frantz Fanon, BP 399, 16000 Alger, Algeria}
\address{\dag Universit\'e de Blida, Institut d'A{\'e}ronautique,
 Blida, Algeria}
\address{\ddag\ LPTL, Universit\'e de Paris 6, Tour 24,
 4\`eme \'etage, case 121, 4 Place
Jussieu, 75252 Paris Cedex 05, France} 
\ead{maillard@lptl.jussieu.fr, njzenine@yahoo.com}

\begin{abstract}
In a previous paper (J. Phys. A {\bf 37} (2004) 9651-9668) we have given the
Fuchsian linear differential equation satisfied by $\chi^{(3)}$,
the ``three-particle'' contribution to the susceptibility of the 
isotropic square
lattice Ising model. This paper gives  the details of
the calculations (with some useful tricks and tools) allowing 
one to obtain long series in polynomial time.
The method is based on series expansion in the variables
 that appear in the $(n-1)$-dimensional integrals representing
the $n$-particle contribution to the isotropic square lattice Ising model
susceptibility $\chi $. The integration rules are straightforward
due to remarkable formulas we derived for these variables. 
We obtain without any numerical approximation  $\chi^{(3)}$ as a fully
integrated series in the
variable $w=s/2/(1+s^{2})$, where $\,s\,=sh (2K)$, with $K=J/kT$ the
 conventional Ising model coupling constant. We also give some
 perspectives and comments on these results.
\end{abstract}

\vskip .5cm


\noindent {\bf AMS Classification scheme numbers}: 34M55, 47E05, 81Qxx, 32G34, 34Lxx, 34Mxx, 14Kxx 


\section{Introduction}
Since the formal expression for the magnetic susceptibility of 
square lattice Ising
model, derived by T.T. Wu, B.M. McCoy, C.A. Tracy and 
E. Barouch~\cite{wu-mc-tr-ba-76}, and the
exact expression of the Ising susceptibility written as an infinite sum: 
\begin{eqnarray}
\label{TTWu}
\chi (T) \, =\, \, \sum_{n=1}^{\infty }\chi ^{(n)}(T)
\end{eqnarray}
of $(n-1)$-dimensional integrals~\cite%
{nappi-78,pal-tra-81,yamada-84,yamada-85,nickel-99,nickel-00}, the
integration of the latter has become a long-standing problem in statistical
physics. The above sum is restricted to odd (respectively even) $n$ for the
high (respectively low) temperature case. While $\chi ^{(1)}$ is obtained
directly without integration, and $\chi ^{(2)}$ is given in terms of
elliptic integrals, \emph{no closed forms} for the higher terms are known.
It is only recently that the differential equation for $\chi^{(3)}$ has
been found \cite{ze-bo-ha-ma-04a}.

Recalling the particle physics terminology of 
T. T. Wu \textit{et al.}, these
$(n-1)$-dimensional integrals are seen as successive \textquotedblleft
particle excitations\textquotedblright , the first one, $\chi ^{(1)}$,
appears to have been calculated by Syozi and Naya~\cite{syo-nay-60} for 
\emph{anisotropic} Ising models (on the triangular lattice), and is actually
a very good approximation to the susceptibility $\chi $, its
{\em anisotropic} series expansion being also 
in remarkable agreement~\cite{HaMa88}
 with the anisotropic series expansion for $\chi $:
\begin{eqnarray}
\label{Naya}
&&\chi ^{(1)}(t_{1},t_{2},t_{3}) = 
 {{ (1-t_1^2)(1-t_2^2) (1-t_3^2)} \over {
(1-t_1 -t_2 -t_3 -t_1 t_2  -t_2 t_3  -t_1 t_3  +t_1 t_2 t_3)^2}} \cdot {M^{*}}^2 , 
 \nonumber   \\
&& M^{*} =  \Bigl( 1 - {{16 (1 +t_1 t_2 t_3) (t_1 + t_2 t_3)
(t_2 + t_1 t_3)(t_3 + t_1 t_2)
} \over {  (1-t_1^2)^2 (1-t_2^2)^2 (1-t_3^2)^2
}} \Bigr)^{1/8} 
\end{eqnarray}
where $\,t_{1}$,  $\,t_{2}$ and $\,t_{3}$ are the three standard high temperature
variables, $\,t_{i}\,=\,th(K_{i})$, for the triangular Ising model, and
where $\,  M^{*}$ is nothing but the magnetization of the triangular 
lattice Ising model
for which the low temperature variables have been replaced 
by the high-temperature ones~\cite{HaMa88}.

For the square lattice Ising model, $\chi^{(1)}$ is obtained as the 
$t_3=0$ limit of (\ref{Naya}) for the anisotropic case, and as the $t_3=0$ and
$t_2=t_1$ limit for the isotropic case. 
The two-particle contribution $\, \chi ^{(2)}$ is also known exactly
and can be expressed in terms of the elliptic 
integrals $\, E$ and $\, K$ ~\cite{wu-mc-tr-ba-76} (see also (\ref{khi2}) below). These are the 
only two exact results on the  $n$-particle contributions $\, \chi ^{(n)}$.

Besides these two exact results on the  $n$-particle
 contributions $\, \chi ^{(n)}$, 
the only two exact ``global'' results known 
for the whole susceptibility $\, \chi$,
 correspond to the exact
expression of the susceptibility satisfying
the so-called \emph{disorder conditions}~\cite{Dhar}, and the exact functional 
relation corresponding to the so-called \emph{inversion relation}~\cite{Bax82} for
the susceptibility~\cite{HaMa88}. 

For one of the four  disorder conditions of the checkerboard Ising model,
$\, t_1 + \, t_2\, t_3\, t_4 \, = \, \, 0$, the susceptibility 
of the checkerboard Ising model reduces to the simple rational expression~\cite{Dhar}:
\begin{eqnarray}
\chi(t_1 + \, t_2\, t_3\, t_4=  0)
 \, = \, \, \, {{(1+t_2) \, (1+t_3)\,  (1+t_4)\, (1+ t_2\, t_3\, t_4)} \over 
{ (1\, - t_2\, t_3)\, (1\, - t_3\, t_4) \, \,(1\, - t_2\, t_4)  
}}
\end{eqnarray}
or, more simply, on the anisotropic triangular lattice ($t_4 \, = \, 1$
limit\footnote[1]{
In this limit divide by a factor 2 to get the susceptibility per site.}):
\begin{eqnarray}
\label{kidisord}
\chi(t_1 + \, t_2\, t_3=0) \, = \, \, \, {{(1+t_2) \, (1+t_3)\, (1+ t_2\, t_3)} \over 
{  (1 - t_3) \, \,(1 - t_2) \, (1 - t_2\, t_3) 
}}
\end{eqnarray}
Recall that Syozi and Naya algebraic expression (\ref{Naya}) for the
anisotropic triangular Ising model reduces to the exact expression (\ref{kidisord}) 
when restricted to the disorder condition  
 $\,t_{1}\,+\,t_{2}\,t_{3}\,=\,0\,$. 
Therefore, one finds
that all the $\,\chi ^{(n)}$'s ($n > 1$) of the anisotropic
triangular\footnote[2]{This is also true
for the four disorder conditions on the
checkerboard Ising model, namely:
$\, \chi ^{(n)}(t_1 + \, t_2\, t_3\, t_4=  0)\, = \, \, \, 0$, $n>1$. }
 Ising model should vanish when restricted to the disorder
condition.

The so-called \emph{inversion relation} \cite{Bax82} for
the susceptibility of the checkerboard Ising model 
and for the Ising model on the triangular lattice, read respectively~\cite{HaMa88,JaMa85}: 
\begin{eqnarray} 
&&\chi (t_{1},\,t_{2}, \, t_3, \, t_4)\,
+\,\chi (-t_1,\,1/t_{2}, \,-t_3, \,  1/t_{4})\,=\,\,0  \\
\label{kiinv}
&&\chi (t_{1},\,t_{2}, \, t_3)\,+\,\chi (-t_{1},\,1/t_{2}, \, -t_{3})\,=\,\,0 
\end{eqnarray}
to be combined with the obvious functional equations associated with the 
geometrical symmetries of the model ($\chi (t_{1}, t_{2}, t_3) =  \chi (t_{2}, t_{1}, t_3)=\cdots$).
It is straightforward to verify that the 
algebraic expression (\ref{Naya}) actually satisfies the previous
inverse functional relation (\ref{kiinv}), 
and deduce, order by order in the $n$-particle expansion (\ref{TTWu}),
that all the  $n$-particle
contributions, $\, \chi^{(n)} $, should also satisfy equations 
 (\ref{kiinv}),
together with the  obvious geometrical symmetries.  

Of course all these ideas ($\chi^{(1)}$ versus exact expressions of the
susceptibility restricted to the disorder conditions, inversion relations
together with geometrical functional equations, 
Kramers-Wannier duality, etc)
can be worked out on the checkerboard Ising model (see
 for instance~\cite{HaMa88,HaMa87,HaMa88Int}). 
This kind of analysis underlines the crucial role played by the 
modulus\footnote{Recall that the
spontaneous magnetization of the 
two-dimensional Ising model reads~\cite{yang,chang}:
$ \, M \, = \, \, (1- \, k^2)^{1/8}$. }
of the elliptic function parameterizing the checkerboard Ising model :
 \begin{eqnarray} 
\label{modulus}
 k^2  = \prod_{i=1}^{4} \, {{ 
t_i^{*} \,  (t_i + t_j  t_k  t_l) \,  (1-t_i^{* 2})
} \over {t_i \,  (t_i^{*} + t_j^{*}   t_k^{*} t_l^{*}) \,  (1-t_i^2)
}} \,\, \rightarrow \,\,  k^2  =  \Bigl({{ 1} \over {sh(2K_1)\cdot sh(2K_2)}}
 \Bigr)^2  
\end{eqnarray}
where the $t_j^*$'s denote the dual variables, $t_j^*=(1-t_j)/(1+t_j)$.
For the \emph{anisotropic} square Ising model, the inverse functional
 relation
simply reads~\cite{HaMa88} ($t_3 \, = \, 0$ limit of (\ref{kiinv})):
\begin{eqnarray}
\label{inverse}
\chi (t_{1},\,t_{2})\,+\,\chi (1/t_{1},\,-t_{2})\,=\,\,0
\end{eqnarray}
which can be combined with the obvious symmetry $\,\chi (t_{1},\,t_{2})\,=\,\chi
(t_{2},\,t_{1})\,$.

 Using these last two exact identities, some re-summed
high-temperature series expansions \cite{ha-ma-oi-ve-87,gut-ent-96} 
were performed for the susceptibility $\chi $ of the anisotropic square 
lattice Ising model~\cite{HaMa88,ha-ma-oi-ve-87,gut-ent-96}, showing 
the occurrence of ($2\,k\,+\,1$)-th root
of unity poles~\cite{gut-ent-96}, $\,t_{2}^{2\,k\,+\,1}\,=\,1$, for the $%
(2\,k\,+\,1)$ particle excitations $\,\chi ^{(2\,k\,+\,1)}$. This enabled
A.J. Guttmann and I. G. Enting \cite{gut-ent-96} to conclude that $\,\chi
(t_{1},\,t_{2})$, as a function of $\,t_{2}$, for $\,t_{1}$ fixed, has a
{\em natural boundary} (the unit circle $\,|t_{2}|\,=\,1$),
 and, thus, cannot be
expressed as a solution of a linear differential equation\footnote[4]{
For instance it cannot be expressed in terms of linear combination 
of special functions of
mathematical physics like elliptic integrals or hypergeometric functions.}
of {\em finite} order $\,N$ with polynomial coefficients of 
 degree $\,d$, (even with $\,N$
and $\,d$ very large). Solutions of such linear differential equations with
polynomial coefficients are called {\em holonomic}
 or {\em D-finite}~\cite{lipshitz-89, zeilberger-90}.
Therefore, the susceptibility $\,\chi $ is a \textquotedblleft truly
transcendental function\textquotedblright\ (in $t_2$, $t_1$ fixed)
 if one takes for the definition
of transcendental\footnote[5]{
For example the Gamma function is a transcendental function: it does not
satisfy a linear differential equation, however it verifies
finite-difference (linear and non-linear) functional equations.} that it is
not holonomic or D-finite.
However, the $\,\chi ^{(n)}$'s,
being multiple integrals of simple holonomic (algebraic!) expressions, they must be holonomic :
\emph{they are actually solutions of linear differential equations}.
To sum up, $\chi $ is {\em not} holonomic, but
it is the sum of an infinite number of holonomic expressions. We have, here, a
situation which is totally reminiscent of that encountered with
{\em Feynman graphs} where all the individual Feynman
 contributions are holonomic,
but the sum of all these contributions is not holonomic (see for instance
all the papers on multi-loop Feynman integrals, nested sums and multiple
polylogarithms, Euler-Zagier sums, and other \textquotedblleft
motives\textquotedblright\ theory~\cite{cartier,Tera,Hoff,Oi}). In this respect, the
analysis of the susceptibility $\chi $ should be seen as the simplest
example of such nested sum calculations, and, beyond the legitimate
interest one has for the Ising model, should be seen as a \textquotedblleft
laboratory\textquotedblright\ for such calculations,  hopefully displaying
the emergence of structures and symmetries. In this respect, $\chi $ being
considered as an infinite sum of holonomic $n-$particle excitations $\,\chi^{(n)}$, 
a better knowledge of the analytical structure of
 the $\,\chi^{(n)}$'s is clearly a way to get a deeper understanding  of the still
mysterious analytic structure of $\chi $ in the complex plane.

Let us make here a few remarks. There are many \textquotedblleft levels of
complexity\textquotedblright\ among transcendental
 functions. Here we mean by "transcendental", non-holonomic,
non-D-finite. As far as $\,\chi$ is concerned,
being transcendental,  but expressible as the infinite sum of  holonomic terms, 
 one could seek  solutions of simple, but 
\emph{non-linear}, differential equations. For instance, it is worth
recalling some particular\footnote[3]{
They have fixed critical points~\cite{ince-56} and are associated with some
monodromy preserving property of some associated Schlesinger system~\cite
{Schles3}.} non-linear differential equations, namely, the so-called \emph{
Painlev\'{e} transcendentals}~\cite{Painl,Schles4} that are
 known to occur for some two-point
correlation functions of the Ising models~\cite{wu-mc-tr-ba-76,Schles2} (Painlev\'{e}
transcendentals do not have natural boundaries). As a consequence
of the previously mentioned natural boundary, the susceptibility $\,\chi $
cannot be a solution of a linear differential equation, but it may well be a
solution of a \emph{non-linear} differential equation, even 
 a very simple one! Actually, it is also worth recalling 
the {\em Chazy equation}~\cite{chazy,chazy2}:
\begin{eqnarray}
\label{chazy}
 {{d^3} \over{dw^3}} y  \, \,=\,\,   \,
 2 \, y \,  {{d^2} \over{dw^2}} y \, - 3\, \Bigl({{d} \over{dw}} y \Bigr)^2
\end{eqnarray}
which arises in the study of third order ordinary 
differential equations having the ``Painlev\'e
property''\footnote[4]{The solutions have only poles for 
movable singularities~\cite{ince-56}. Movable means that the singularity's
position varies as a function of the initial values.} ~\cite{Painl2}.
This non-linear equation
is probably the simplest example of an ordinary differential
equation whose solutions {\em have a  natural
boundary}. For the Chazy equation (\ref{chazy}), the
 boundary is a {\em movable circle}\footnote[5]{Heuristically, this can be simply understood 
introducing exponentially small terms in the expansion of the solutions
(see (3.37) in~\cite{Kruskal}), or as Chazy did~\cite{chazy,chazy2}, by 
relating its solutions to those of a linear hypergeometric equation, 
performing an infinite process of analytic continuations
yielding a tesselation of the interior (respectively exterior) of a circle.
} (its position depends on the initial conditions). The whole susceptibility $\,\chi $, being
 the sum of all the two-point correlation
functions, one could, thus, perfectly imagine that $\,\chi $, {\em despite 
its unit circle natural boundary}~\footnote[1]{The susceptibility $\,\chi $ 
of the {\em isotropic} Ising model has 
a unit circle natural boundary in the variable
 $\,s \, = \, sh(2\, K)$, see (\ref{nick}) below
and ~\cite{nickel-99,nickel-00}.}, could 
well be a solution of a simple
{\em non-linear} differential equation related to 
 Painlev\'{e}-like transcendants.

While there are {\em no algorithms} to find such {\em non-linear}
differential equations, implementing an algorithm to 
seek for a linear differential equation
reduces to solving a set of linear equations. 
One knows that each of the $\,\chi^{(n)}$'s is a solution of
a linear differential equation, however, and unfortunately, the
mathematical theorems that prove that the $\,\chi ^{(n)}$'s are holonomic,
because they are multiple integrals of holonomic expressions, \emph{do not
give any upper bound on the order of the linear differential equation and no
upper bound of the degree of the polynomials in front of the successive
derivatives either}~\cite{lipshitz-89,zeilberger-90}.
One can try to get enough coefficients of some high
(respectively low) temperature series expansion to actually
\textquotedblleft guess\textquotedblright\ the linear differential
equation satisfied by a given  $\,\chi^{(n)}$.
Recalling the previous exact results that arise from a study of the  anisotropic
model, it might be tempting to generate {\em anisotropic} high temperature series
expansions in order to use these exact results to get severe exact
constraints on the coefficients of the series expansion.
This can be done in principle, however, the number of
coefficients in the two variables, and the combinatorial complexity
 associated with the finding of these coefficients, is too large compared to
the \textquotedblleft advantage\textquotedblright\ one gets from the previous exact
constraints. Therefore, we will ignore the previous exact anisotropic
results, and  will generate more standard
{\em isotropic} series expansions for $\,\chi^{(n)}$, 
and more specifically for $%
\,\chi^{(3)}$.

Recall that a huge amount of work had
 already been performed by B. Nickel~\cite%
{nickel-99,nickel-00} to
generate isotropic series coefficients for $\chi ^{(n)}$. More
 recently one should mention the generation by W.
Orrick \textit{et al.}~\cite{or-ni-gu-pe-01b}, of coefficients of $\chi $
using  \emph{non-linear Painlev\'{e} difference 
equations}\footnote[8]{More generically
this corresponds to the notion of the so-called
 Hirota bilinear equations.} for the
correlation function~\cite%
{or-ni-gu-pe-01b,or-ni-gu-pe-01,coy-wu-80,perk-80,jim-miw-80}. This second
method, because it uses an exact non-linear difference equation was able to
provide an algorithm for computing the successive coefficients in  \emph{%
polynomial} time~\cite{or-ni-gu-pe-01b} (namely $\,O(N^{6})$), instead of the
exponential growth one expects at first sight. Roughly speaking, the first
calculations by B. Nickel~\cite{nickel-99,nickel-00} were actually performed
using the multiple integral form of the $\,\chi ^{(n)}$'s and were thus able
to provide series expansions for the $\,\chi ^{(n)}$'s {\em separately}. In contrast, the
second method, which takes into account a fundamental non-linear symmetry,
namely non-linear Painlev\'{e} difference
 equations~\cite{coy-wu-80,perk-80,jim-miw-80},
provides, as a consequence, a series expansion for the 
whole susceptibility $\,\chi $, where the
various $\,\chi ^{(n)}$'s are difficult to disentangle. To sum up roughly,
the first standard method is  holonomic, or linear, oriented
(decomposition in the holonomic $\,\chi ^{(n)}$'s) while the second method
is more non-holonomic, \textquotedblleft non-linear
oriented\textquotedblright\ (transcendental Painlev\'{e} symmetries ...).

As far as the singular behavior near the dominant ferromagnetic point
$\,T\,=\,T_{c}$ is concerned, it is numerically extremely hard to disentangle
the $\,|T-T_{c}|^{9/4}$, $\,|T-T_{c}|^{9/4}\cdot \log|T-T_{c}|$ and
$\,|T-T_{c}|^{2}\cdot \log|T-T_{c}|$ contributions, and this was certainly a
motivation to get longer series. This set of extensive results~\cite
{nickel-99,nickel-00,or-ni-gu-pe-01b} (expansion up to order 116, then to
order 257, and to order 323 for $\chi $) already enabled one to get a better
understanding of the various singular behaviors. The short-distance terms
were shown to have the form $\,(T-T_{c})^{p}\cdot (\log|T-T_{c}|)^{q}\,$ with 
$\,p\geq q^{2}$~\cite{or-ni-gu-pe-01b, or-ni-gu-pe-01}.
As far as the other singular points are concerned (non physical
singularities in the complex plane), B. Nickel~\cite{nickel-99}, in a
remarkable paper showed that $\,\chi ^{(2\,n+1)}$ is singular for the
following values of $\,s=sh (2J/kT)$ lying on the \emph{unit circle}
($k=m=0$ excluded): 
\begin{eqnarray}
\label{nick}
&&
2\cdot (s\,+\,{\frac{{1}}{{s}}})\,=\,\,u^{k}\,+\,{\frac{{1}}{{u^{k}}}}%
\,+\,u^{m}\,+\,{\frac{{1}}{{u^{m}}}}\,    \label{sols} \\
&&\qquad u^{2\,n+1}\,=\,\,1,\qquad \qquad -n\,\,\leq \,\,m,\,\,k\,\,\,\leq
\,\,n \nonumber
\end{eqnarray}
the singularities being logarithmic branch points of order
 $\,\epsilon^{2\,n\,(n+1)-1}\cdot \ln(\epsilon )$ with $\,\epsilon \,=\,1-s/s_{i}$ where $%
\,s_{i}$ is one of the solutions of (\ref{sols}). This confirms, this
time for the \emph{isotropic model} the existence of a natural boundary for $%
\,\chi (s)$, namely $\,|s|\,=\,1$, and that $\,\chi (s)$ is a \emph{%
transcendental} function: it is not D-finite (holonomic) as a function of $%
\,s$.

Our approach here is clearly of holonomic type,
underlining the crucial role played by some well-suited {\em hypergeometric}
functions. The emergence of these structures actually allows us to get
extremely efficient \emph{polynomial time} calculations
(namely $\,N^{4}$, for $\chi^{(3)}$),
which enable us, in particular, to calculate the series coefficients
\emph{separately},
and \emph{not from a recursion} that requires the storage of all the
previous data. In the sequel, we give a fully integrated expansion of the
three-particle susceptibility contribution $\,\chi^{(3)}$
as {\em multisums of hypergeometric functions}.
This expression is used to generate a long series in the variable
$w=s/2/(1+s^{2})$. With this long series, we
 succeeded~\cite{ze-bo-ha-ma-04a} in obtaining the
homogeneous seventh order Fuchsian differential 
equation satisfied by $\chi^{(3)}$.

This paper is organized as follows: in Section 2, we present the basic
features of our expansion method that allow us to obtain the fully integrated $%
\,\chi ^{(3)}$ as a multisum of products of three hypergeometric functions,
without any numerical approximation. The three-particle susceptibility
 $\,\chi^{(3)}$ is written, for series generation
purposes, as the sum of a closed, but involved, expression (sum of
linear, quadratic and cubic products of elliptic integrals), of a second
simple (non closed, for the moment, but easy enough to compute) sum, and a
last sum which is much harder to calculate, and which requires most of the computing time.
The variable $\, w$ being well-suited to both high and low temperature
regimes, the expansion method we use is 
generalizable to other $\chi^{(n)}$'s ($\, n$ odd, or even).
In Section 3, we recall the homogeneous linear differential equation and
discuss the new singularities discovered in \cite{ze-bo-ha-ma-04a}.
In Section 4, using a differential approximant method, we show how the new
non apparent singularities can be discovered even when the series are not
long enough to obtain the exact linear differential equation.
Finally, Section 5 contains our conclusions.

\section{The fully integrated expansion of $\protect\chi^{(3)}$ }
\label{expchi3}
\subsection{The expansion method}
Let us focus on the third contribution to the susceptibility $\chi$
defined by the double integral as given in \cite{nickel-99}:
\begin{eqnarray}
\chi ^{(3)}(s) &=&(1-s^{4})^{1/4}/s\,\cdot \,\tilde{\chi}^{(3)}(s) 
\nonumber
\label{chi3tild} \\
\tilde{\chi}^{(3)}(s) &=&{\frac{1}{4\pi ^{2}}}\int_{0}^{2\pi }d\phi
_{1}\int_{0}^{2\pi }d\phi _{2}\,\cdot \tilde{y}_{1}\tilde{y}_{2}\tilde{y}%
_{3}\cdot \Bigl({\frac{1+\tilde{x}_{1}\tilde{x}_{2}\tilde{x}_{3}}{1-\tilde{x}%
_{1}\tilde{x}_{2}\tilde{x}_{3}}}\Bigr)\cdot H^{(3)} 
\end{eqnarray}
with 
\begin{eqnarray}
\tilde{x}_{j} &=&{\frac{s}{1+s^{2}-s\cos {\phi _{j}}+\sqrt{(1+s^{2}-s\cos {%
\phi _{j}})^{2}-s^{2}}}},  \label{varxy} \\
\tilde{y}_{j} &=&{\frac{s}{\sqrt{(1+s^{2}-s\cos {\phi _{j}})^{2}-s^{2}}}}%
,\quad  j=1,2,3, \quad \phi _{1}+\phi _{2}+\phi _{3}=0.  \nonumber
\end{eqnarray}
Many forms for $H^{(3)}$ may be taken \cite{nickel-99,nickel-00} and are
equivalent for integration purposes, e.g., 
\begin{eqnarray}
\label{h3old}
H^{(3)}\,=\, f_{23} \cdot \Bigr(f_{31}+{\frac{{f_{23}}}{{2}}}\Bigl),\quad 
f_{ij}\,=\,(\sin {\phi _{i}}-\sin {\phi _{j}})\cdot {\frac{\tilde{x}_{i}\tilde{x}%
_{j}}{1-\tilde{x}_{i}\tilde{x}_{j}}}. 
\end{eqnarray}

Instead of the variable $s$, we found it more suitable to use $w={\frac{{1}}{{2}%
}}s/(1+s^{2})$ where the unit circle $ | s | =1$
 becomes $] -\infty, -1/4 ] \cup [ 1/4, \infty [ $ in $w$. 
This variable allows to deal with both limits (high and low $%
s$) on an equal footing.

From now on, we will use the scaled variables:
\begin{eqnarray}
x_{j} &=&\frac{\tilde{x}_{j}}{w}={\frac{2}{1-2w\cos {\phi _{j}}+\sqrt{%
(1-2w\cos {\phi _{j}})^{2}-4w^{2}}}},  \label{varxy2} \\
y_{j} &=&\frac{\tilde{y}_{j}}{2w}={\frac{1}{\sqrt{(1-2w\cos {\phi _{j}}%
)^{2}-4w^{2}}}} \nonumber
\end{eqnarray}
which behave like: $1+O(w)$ at small $w$. These variables are related by:
\begin{eqnarray}
\label{yx}
y_j \cdot (1-w^2 \, x_j^2) \, = \, x_j.
\end{eqnarray}

Performing the integrals in (\ref{chi3tild}) is highly non trivial. For this,
one may expand symbolically the integrand in (\ref{chi3tild}) in the variable $w$, and integrate
the angular part. This way one faces some 18 sums to carry out. These sums
come from the expansion of the quantities $x_{j}$ and $y_{j}$, and also of the
denominators. The convoluted character of the integrand, and especially $%
1/(1-w^{3}x_{1}x_{2}x_{3})$, makes the symbolic expansion in $w$ almost intractable.

To overcome this difficulty we expand
the integrand of $\,\tilde{\chi}^{(3)}$ in the various
variables $x_{j}$. The expansion depends now only on combinations of the
form :
\begin{eqnarray}
\label{formeint}
y_{1}x_{1}^{n_{1}}\cdot y_{2}x_{2}^{n_{2}}\cdot y_{3}x_{3}^{n_{3}}\cdot
P_{n_{1},n_{2},n_{3}}(\phi _{1},\phi _{2},\phi _{3})  
\end{eqnarray}
where $P_{n_{1},n_{2},n_{3}}(\phi _{1},\phi _{2},\phi _{3})$ is a
polynomial in circular functions of the angles $\phi _{j}$. 
The structure (\ref{formeint}) 
appearing in the integrand, after the expansion in the variables 
$x_{j}$, is independent of the particular form taken for $H^{(3)}$.

We succeeded in deriving a remarkable formula for $y_{j}x_{j}^{n}$ that
carries only {\em one summation index} and reads (we drop the indices): 
\begin{eqnarray}
\label{FourExp}
yx^{n}\,=\,\,a(0,n)\,+2\sum_{k=1}^{\infty }w^{k}\,a(k,n)\,\cos {k\phi }
\end{eqnarray}
where $a(k,n)$ is a {\em non terminating} hypergeometric series that reads: 
\begin{eqnarray}
\label{defa}
&& a(k,n) =  {m \choose k}\cdot \\
&& {_{4}}F_{3}\Bigl({\frac{{(1+m)}}{{2}}},
{\frac{{(1+m)}}{{2}}},{\frac{{(2+m)}}{{2}}},{\frac{{(2+m)}}{{2}}};
1+k,1+n,1+m;16w^{2} \Bigr)  \nonumber
\end{eqnarray}
where $m=k+n$. Note that $a(k,n)=a(n,k)$.

With this Fourier series, the angular integration of the form (\ref{formeint}%
) becomes straightforward\footnote[2]{%
Note that we can deduce from the Fourier series (\ref{FourExp}) that $%
yx^{n}$ also has an expansion in the variable $w$ (respectively $\sin ^{2}(\phi /2)$%
) with coefficients as {\em hypergeometric functions} of the variable $\sin
^{2}(\phi /2)$ (respectively $w$). Actually, the expansion (\ref{FourExp}) is more
suitable for angular integration purposes.}.

The method described above being independent of the particular choice of $%
H^{(3)}$, let us consider the form (\ref{h3old}). Noting that :
\begin{eqnarray}
{\frac{\tilde{x}_{i}\tilde{x}_{j}}{1-\tilde{x}_{i}\tilde{x}_{j}}}\,\,=\,\,\,{%
\frac{w}{2}}\,{\frac{x_{i}-x_{j}}{\cos \phi _{i}-\cos \phi _{j}}}
\end{eqnarray}
the $f_{ij}$'s can be rewritten as : 
\begin{eqnarray}
f_{ij}\,=\,\,-{\frac{w}{2}}\,(x_{i}-x_{j})\,\cot (\phi _{i}/2+\phi _{j}/2).
\end{eqnarray}%
$H^{(3)}$ becomes simply a polynomial in the $x_{i}$'s : 
\begin{eqnarray}
\label{h3new}
&& H^{(3)} = {\frac{w^{2}}{4}}\cdot \\
&& \Bigl( {{1}\over{2}} (x_{2}-x_{3})^2 \cdot \cot ^{2}({{\phi_{1}}\over{2}})-
(x_{1}-x_{3})(x_{2}-x_{3})\cdot \cot ({{\phi_{1}}\over{2}})\cdot
\cot ({{\phi_{2}}\over{2}})\Bigr).  \nonumber
\end{eqnarray}

Expanding the integrand of $\tilde{\chi}^{(3)}$ in the variables $x_{j}$ :  
\begin{eqnarray}
\label{khi3tilda}
\tilde{\chi}^{(3)}(w)  &=&  {\frac{8w^{3}}{4\pi ^{2}}}\int_{0}^{2\pi }d\phi
_{1}\int_{0}^{2\pi }d\phi _{2}\cdot \\
&& \quad y_{1}y_{2}y_{3}\cdot
\Bigl(1+2\sum_{n=1}^{\infty }w^{3n}(x_{1}x_{2}x_{3})^{n}\Bigr)\cdot H^{(3)}
\nonumber
\end{eqnarray}
together with the expression (\ref{h3new}) for $H^{(3)}$, one notes that
the integrand depends only on combinations of the form :   
\begin{eqnarray}
&&y_{1}x_{1}^{n_{1}}\cdot y_{2}x_{2}^{n_{2}}\cdot y_{3}x_{3}^{n_{3}}\cdot
\cot ^{2}(\phi _{1}/2) \\
&&y_{1}x_{1}^{n_{1}}\cdot y_{2}x_{2}^{n_{2}}\cdot y_{3}x_{3}^{n_{3}}\cdot
\cot (\phi _{1}/2)\cot (\phi _{2}/2)
\end{eqnarray}
which have straightforward integration rules. 

The problem of integration is thus settled with a 
{\em limited number of sums}. In Appendix A we give the
integration rules used and show how the cancellation of the ``artificial''
singularity introduced by taking the particular form (\ref{h3new}) for $%
H^{(3)}$ occurs.

The expansion of the integrand in the variables $x_{j}$, together with the
remarkable formula (\ref{FourExp}), allow us to obtain $\tilde{\chi}^{(3)}(w)$ as
a fully integrated expansion (see Appendices B and C), namely: 
\begin{eqnarray}
 \label{khi3sergen}
\tilde{\chi}^{(3)}(w)\,=\,\,8\ w^{9}\cdot \Bigl(\tilde{\chi}_{cl}^{(3)}(w)\,
+\, 2\, w^{4}\cdot \Xi(w) 
\Bigr)  
\end{eqnarray}
where $\Xi(w) $ are sums given in Appendix B, and $\tilde{\chi}_{cl}^{(3)}(w)$
is a {\em closed expression}: 
\begin{eqnarray}
\label{clos}
&& \tilde{\chi}_{cl}^{(3)}(w)= \, {\frac{1}{256\,(1-4w)^{3/2}w^{13}}} \cdot \\ 
&& \quad \Bigl(\sqrt{1-4w}\sum_{i=0}^{3}
\sum_{j=0}^{i}q_{j,i-j}\tilde{K}^{j}\tilde{E}
^{i-j}+\sum_{i=0}^{2}\sum_{j=0}^{i}p_{j,i-j}\tilde{K}^{j}\tilde{E}^{i-j}%
\Bigr)  \label{khi3closed} \nonumber 
\end{eqnarray}
$\tilde{K}$ and $\tilde{E}$ being functions of $w$ related to elliptic
integrals $K$ and $E$:
\begin{eqnarray}
\label{KE}
&&\tilde{K}\,=\, \, {_{2}}F_{1}\Bigl(1/2,1/2;1;16w^{2}\Bigr)=\,2\,K(4\,w)/\pi
,\qquad   \nonumber   \\
&&\tilde{E}\,=\, \, {_{2}}F_{1}\Bigl(-1/2,1/2;1;16w^{2}\Bigr)=\,2\,E(4\,w)/\pi ,
\end{eqnarray}
and where the $q_{i,j}$'s and $p_{i,j}$'s are polynomials in $w$ given in Appendix D.
One should note that this closed expression (\ref{clos}) is not "natural":
the separation in (\ref{khi3sergen}) is not unique.
It comes from our ability to evaluate some sums,
and is only made for series generation purposes
(see Appendices B and C for details).
In fact, $\tilde{\chi}^{(3)}$ in (\ref{khi3sergen}) can be written as a
multisum of product of three hypergeometric functions $a(k,n)$,
as can be clearly seen\footnote[3]{For $\Xi(w)$ this multisum
of product of three hypergeometric functions can be seen in the
definition of the auxillary functions (see Appendices A, B and C)
in terms of $a(k,n)$.} in (\ref{formeint}).

One recalls that the key ingredient in obtaining (\ref{khi3sergen})
 is the Fourier expansion of $y\,x^{n}$, a quantity appearing
in any $\chi ^{(n)}$. The method is thus generalizable,
independent of whether $n$ is even or odd, since the variable $w$ has, by
construction, Kramers-Wannier duality invariance 
($s \,  \leftrightarrow \, 1/s$).
Appendices E and F show how the known result~\cite{wu-mc-tr-ba-76}
for $\chi^{(2)}$ appears, in this framework, as a sum of products of two
hypergeometric functions merging 
into a {\em single} hypergeometric function.

For the $\tilde{\chi}^{(3)}$ case, we see
that the  mechanism of {\em fusion
relations} on hypergeometric functions that we saw previously
 for $\tilde{\chi}^{(2)}$,
may also hold similarly. However, we have not discovered a similar fusion
 mechanism for $\tilde{\chi}^{(3)}$ .

\subsection{Series generation}
From our integrated forms of $\tilde{\chi}^{(3)}$, the generation of series
coefficients becomes straightforward. \emph{Recall that} (\ref%
{khi3sergen}) \emph{ is already integrated}, and, thus, the computing time to obtain
the series coefficients comes from the evaluation of the sums.
For the forms used, this time\footnote[3]{
The overall computing time can be reduced by using the
inhomogeneous recurrences on, and relations between the coefficients of the
functions introduced (see Appendix G).
} is of order $N^{4}$. 

We have been able to generate a long series of coefficients from the
expression (\ref{khi3sergen}) up to order 490. 
More precisely, the series expansion for  $\Xi $ in (\ref{khi3sergen}),
was, in fact, obtained
as the sum of two series, a ``hard to compute'' series up to order 490, $\Xi_{h} $,
requiring most of the computing
time\footnote[4]{Expression $\Xi_{h} $ actually corresponds to the true limitation
in our series expansion. It corresponds to the third term at the right
hand side of (\ref{Xi}), see Appendix B below.},
and a simpler one, $\Xi_{s} $,  requiring much less computing
 time, but of course, more time than the closed
expression $\, \tilde{\chi}_{cl}^{(3)}$. 
They read respectively :
\begin{eqnarray}
\label{expandxih}
2 w^4 \cdot \Xi_{h}(w)  = 
16{w}^{4}+88{w}^{5}+1008{w}^{6}+5144{w}^{7}
 + \cdots 
\end{eqnarray}
\begin{eqnarray}
\label{expandxis}
2 w^4 \cdot \Xi_{s}(w)  = 
4{w}^{4}-4{w}^{5}+6{w}^{6}-1628{w}^{7}-11738{w}^{8} + \cdots 
\end{eqnarray}
and
\footnote[5]{
This sum is actually analytic near $\, w \, = \, 0$,  
while the expansions of the various cubic sums, quadratic sums and
linear sums in  $K$ and $E$ in the closed form (\ref{clos})
exhibit $1/w^{13}, \,1/w^{12},  \cdots $ poles.} :
\begin{eqnarray}
\label{closexp}
\tilde{\chi}_{cl}^{(3)}(w) =  1+36{w}^{2}+4{w}^{3}
+864{w}^{4}+112{w}^{5}+17518{w}^{6}
 +  \cdots 
\end{eqnarray}
The expansion of $\tilde{\chi}^{(3)}$ thus reads : 
\begin{eqnarray}
\label{series}
&&{\frac{\tilde{\chi}^{(3)}(w)}{8\, w^{9}}} =
1+36{w}^{2}+4{w}^{3}+884{w}^{4}+196{w}^{5}+18532w^6+\cdots   
\\
&& 4309904156072449034005168913683182133673083305758179058 \nonumber \\
&& 1338044228506645721942836227131943916427611183965848882 \nonumber \\
&& 0854328108563517776050668285313206447166513353732686903 \nonumber \\
&& 6438360433784746161733072061779435657865203525518535003 \nonumber \\
&& 4549334011001139634569887672761933044541359314192379119 \nonumber \\
&& 5381220947969620\cdot w^{481}\,+\,\cdots   \nonumber
\end{eqnarray}
in agreement with the coefficients obtained by B. Nickel \cite{nickel-00}
and those obtained by A. J. Guttmann \textit{et al.}\footnote[1]{%
Private communication : W. Orrick and A. J. Guttmann obtained 
the series up to 250 coefficients, in agreement with our 
490 coefficients.}, these last expansions being, in fact, written in terms of 
 the $\,u \, = \,s/2$ variable~\cite{nickel-00} :
\begin{eqnarray}
\label{seriesU}
{{\tilde{\chi}^{(3)}(u)} \over {4 u^8}} 
 =  1+4{u}^{3}+16{u}^{4}+4{u}^{5}+20{u}^{6}
+84{u}^{7}+247{u}^{8} 
 + \cdots 
\end{eqnarray}
For large values of $N$, the coefficients $C_{N}$, on the 
right hand side of (\ref{series}) and (\ref{seriesU}), grow, 
in the variable $w$ as:
\begin{eqnarray}
C_N \simeq 13.5\times 4^{N},  \quad N\, {\rm even}, \qquad 
C_N \simeq 11\times 4^{N}, \quad  N\, {\rm odd}
\end{eqnarray}
and in the variable $\, u=\, s/2$,  as
$\, C^{(u)}_N \simeq 15\times 2^{N}$.

Our long series is obtained as the sum of three contributions of
different algorithmic complexity. The contribution (\ref{closexp}) being
given by an exact closed form, 
we were tempted to seek a linear differential equation satisfied by
$\tilde{\chi}_{cl}^{(3)}$. Appendix H contains a report on this
matter, which suggests that seeking for the differential equation for
$\tilde{\chi}^{(3)}$, seen as the sum (\ref{khi3sergen}),
is in fact much simpler than seeking the differential equation for
each constituent. The whole sum (\ref{khi3sergen}) is, thus, a better
candidate for seeking for a linear differential equation.

\section{The Fuchsian differential equation satisfied by $\tilde{\protect\chi%
}^{(3)}$}
\label{fuchsdiff}
With our long series, and with a dedicated program, we have succeeded in obtaining
the differential equation for $\, \tilde{\chi}^{(3)}$ that is
given in \cite{ze-bo-ha-ma-04a} and recalled here:
\begin{eqnarray}
\sum_{n=0}^{7}\,a_{n}(w)\cdot {\frac{{d^{n}}}{{dw^{n}}}}F(w)\,\,=\,\,\,\,\,0
\label{fuchs}
\end{eqnarray}
with : 
\begin{eqnarray} 
\label{defQ}
&&a_{7}=w^{7} \left( 1-w\right) \left( 1+2w\right) {\left(
1-4w\right) }^{5}{\left( 1+4w\right) }^{3}\left(
1+3w+4w^{2}\right) P_{7}(w),  \nonumber \\
&&a_{6}=w^{6} {\left( 1-4w\right) }^{4}{\left( 1+4w\right) }^{2}P_{6}(w),
\,\, a_{5}= w^{5} {\left( 1-4w\right) }^{3}\left( 1+4w\right)P_{5}(w),  \nonumber \\
&&a_{4}=w^{4} \left(1-4w\right) ^{2}P_{4}(w),\qquad \quad \quad
a_{3}= w^{3}  \left( 1-4w\right) P_{3}(w),  \nonumber \\
&&a_{2}=w^{2}  P_{2}(w),\qquad \quad a_{1}=w  P_{1}(w),\qquad
\quad a_{0}=P_{0}(w).  
\end{eqnarray}
where $P_{7}(w),P_{6}(w)$ $\cdots $, $P_{0}(w)$ are polynomials of degree
respectively 28, 34, 36, 38, 39, 40, 40 and 36 given in \cite{ze-bo-ha-ma-04a}.

The differential equation (\ref{fuchs}) is an equation of the \emph{Fuchsian
type since there are no singular points, finite or infinite, other than
regular singular points}. The singularities correspond to the roots of the
polynomial in front of the highest derivative in (\ref{fuchs}).
The ferromagnetic ($w=1/4$ or $s=1$),
antiferromagnetic ($w=-1/4$ or $s=-1$), zero or infinite temperature ($w=0$
or $s=0,\infty $), non-physical ($w=\infty $ or $s=\pm i$) and Nickel
singularities ($w=1,\,-1/2$ or $2-s+2s^{2}=0,\,1+s+s^{2}=0$) are indeed
regular singular points of the differential equation. The last two are
given by Nickel's relation (\ref{nick}) 
for $\,n=1\,$. Besides the known singularities mentioned 
above, we note the occurrence of
the roots of the polynomial $\,P_{7}(w)$ of degree 28 in $w$, and of the two \emph{%
quadratic numbers} $\,1+3\,w+4\,w^{2}\,=\,0$ which {\em are  
not} of Nickel's form (\ref{nick}).

Applying Frobenius's method \cite{ince-56}, we give in Table 1, for each
regular singular point, the critical exponents, 
the number of solutions in Frobenius form, and the maximum power of the
logarithmic terms.
One sees that only $\log$ and $\log^2$ terms appear in the local solutions
of (\ref{fuchs}).

\textbf{Table 1:} Critical exponents for each regular singular point. 
$N_{F}$ and $P$ are, respectively, the
number of solutions in Frobenius form, and the maximum power of the logarithmic
terms for each singularity of (\ref{fuchs}).
$w_{P}$
is any of the 28 roots of $P_{7}(w)$. We have also shown the corresponding
roots in the $s$ variable. In the variable $\,s$ the local exponents for
$w=\pm 1/4$ are twice those given.
\centerline{
\begin{tabular}{|l|l|l|l|l|}
\hline
&  &  &  &  \\ 
$w$-singularity & $s$-singularity & Critical exponents in $w$ & $N_{F}$ & $P$
\\ 
&  &  &  &  \\ 
\hline
&  &  &  &  \\
$0$ & $0,\infty$ & $9,3,2,2,1,1,1$ & $4$ & $2$ \\ 
&  &  &  &  \\ 
$-1/4$ & $-1$ & $3,2,1,0,0,0,-1/2$ & $4$ & $2$ \\ 
&  &  &  &  \\ 
$1/4$ & $1$ & $1,0,0,0,-1,-1,-3/2$ & $4$ & $2$ \\ 
&  &  &  &  \\ 
$-1/2$ &  $\frac{-1\pm i\sqrt{3}}{2}$   & $5,4,3,3,2,1,0$ & $6$ & $1$ \\ 
&  &  &  &  \\ 
$1$ & $\frac{1\pm i\sqrt{15}}{4}$  & $5,4,3,3,2,1,0$ & $6$ & $1$ \\ 
&  &  &  &  \\ 
$\frac{-3\pm i\sqrt{7}}{8}$ & $\frac{-1\pm i\sqrt{7}}{4}$, $\frac{-1\pm i\sqrt{7}}{2}$  & $5,4,3,2,1,1,0$
& $6$ & $1$ \\ 
&  &  &  &  \\ 
$\infty $ & $\pm i$ & $3,2,1,1,1,0,0$ & $4$ & $2$ \\ 
&  &  &  &  \\ 
$w_{P}$, 28 roots & $s_{P}$, 56 roots & $7,5,4,3,2,1,0$ & $7$ & $0$ \\ 
&  &  &  &  \\ \hline
\end{tabular}
}

\vskip 0.5cm

One can see that the roots of 
$P_7$ are {\em apparent singularities}, the 
local solutions carry {\em no logarithmic terms}  
and {\em are analytic} since the exponents {\em are all positive
integers}.
Consider the apparent singularity effect from the series expansion
$\tilde{\chi}^{(3)}$ viewpoint of this paper.
If the roots of $P_7$ were not apparent and were singularities of $\tilde{\chi}^{(3)}$,
the coefficients of (\ref{series}), instead 
of being dominated by the $\, w \, = \, \pm 1/4$ singularities, yielding a
 $\, 4^N$ growth of the coefficients, would be dominated by the smallest
 (in modulus) root of $P_7$, namely  $\, w \, \simeq  \, -0.0424$,
 yielding a growth $\, \simeq \, (23.553)^N \,$ of the coefficients.
We have checked this on many (slight) deformations of (\ref{fuchs}), keeping
all but one of the $\, P_n$'s unchanged (in particular $P_7$), and slightly modifying
 one $P_n$ ($n \, \ne \, 7$). One immediately sees a
 {\em change in the growth of the coefficients, with the} $\, 4^N$ {\em growth
changing drastically into a} $\, \simeq \, (23.553)^N \,$ {\em growth}.
It is thus crucial that the roots of $\, P_7$ are
 {\em apparent singularities}, which require 
some very specific ``alchemical tuning'' between the various $\, P_n$'s. 
In other words, the Fuchsian equation (\ref{fuchs}) corresponds
 to a rather {\em rigid structure}, since the 
apparent singularity character of $\, P_7$ is an {\em unstable property}.

The two quadratic numbers,
$w \, = \, (-3 \pm i \sqrt{7})/8$ (or $1+3w+4w^2=0$), which read in
the $s$ variable 
\begin{eqnarray}
(2s^2+s+1)(s^2+s+2)\, =\, \, 0
\end{eqnarray}
have their roots lying respectively on the circles $\vert s \vert=1/\sqrt{2}$
and $\vert s \vert=\sqrt{2}$. Near these points, one of the solutions of (%
\ref{fuchs}) has  weak singular behavior (a log singularity). Note however,
 that some of these logarithmic terms, occurring in the general
solution of (\ref{fuchs}) and shown in
 Table 1 {\em may not exist}\footnote[5]{See,
for instance, the two simple rational and algebraic solutions
$\, S_1$ and $\, S_2$ below (see (\ref{S1S2}))
 which are free of the $\, 1-w\, = \, 0$, $\, 1\, +2\, w\, = \, 0$, 
$\, w\, = \, 0$ and $\, 1\, + 3\, w \, + 4\, w^2\, = \, 0$ singularities.}
 in the particular solution $\, \tilde{\chi}^{(3)}$. Actually, the two 
unexpected quadratic numbers correspond to
singularities of solutions of (\ref{fuchs}) which behave locally like
$(1+3 w\, +4\, w^2) \cdot \ln(1+3 w\, +4\, w^2) \cdot q_1\, + \, q_2$,
where $\, q_1$ and $\, q_2$ are analytic 
functions near the two quadratic roots 
$\, 1+3 w\, +4\, w^2\, = \, 0$. This weak singularity\footnote[1]{This ``weakly''
 singular behavior can also be seen in
the monodromy matrix (see~\cite{ze-bo-ha-ma-04a}) 
associated with these two roots,
where one finds a nilpotent matrix of order 
two (no $\, \log^2\, $ term). } gives 
a $\, 2^N$ contribution to the growth of the $w$-coefficients.
However this {\em does not mean} 
that each solution\footnote[2]{A solution of (\ref{fuchs}) can 
exhibit {\em only a restricted set of the whole set of singularities}
of the Fuchsian equation: this is very
 clear with $\, S_1$, or $\, S_2$ (see (\ref{S1S2}) below).} 
  of  (\ref{fuchs}) has such  
$\, (1+3 w\, +4\, w^2) \cdot \ln(1+3 w\, +4\, w^2)$
singular behavior near the two roots $\, 1+3 w\, +4\, w^2\, = \, 0$.
Actually, as far as the ``physical'' solution $\tilde{\chi}^{(3)}$ is
concerned, and as
 far as the series analysis view-point developed in this paper
is concerned, 
this $\, (1+3 w\, +4\, w^2) \cdot \ln(1+3 w\, +4\, w^2)$
singular behavior should be excluded : in the $\, s$ variable,
 instead of the $\, w$ variable, it would give (beyond other terms) a
 $\, (1+s\, +2\, s^2) \cdot \ln(1+s\, +2\, s^2)\, \, $ term,
 yielding a $\, \sqrt{2}^N \,$ 
growth, which would be in contradiction with the
 growth of the coefficients of $\tilde{\chi}^{(3)}$
in the $\, s$ variable.
 This requires further
investigations that we will address in a forthcoming publication.

To end this section and to be complete, we will recall the pertinent
algebraic properties of the Fuchsian differential equation whose analysis
was sketched in \cite{ze-bo-ha-ma-04a}.
The Fuchsian differential equation (\ref{fuchs}) has
two remarkable rational and algebraic solutions, namely: 
\begin{eqnarray}
\label{S1S2}
S_1(w) \, = \, \, {\frac{{w} }{{1\, -4\, w}}}, \qquad \quad S_2(w) \, = \,
\, {\frac {{w}^{2}}{ \left( 1-4\,w \right) \sqrt {1-16\,{w}^{2}}}}
\end{eqnarray}

This results in very important factorization properties for $\,L_{7}$, the
seventh order linear differential operator, corresponding to the Fuchsian
differential equation (\ref{fuchs}): 
\begin{eqnarray}
L_{7}\,=\,\, {\frac{{d^{7}}}{{dw^{7}}}}\,+\,{\frac{1}{a_{7}}}\cdot
\sum_{k=0}^{6}a_{k}\,{\frac{{d^{k}}}{{dw^{k}}}}  \label{defL7}
\end{eqnarray}
The adjoint differential operator $L_{7}^{\ast }$ has the following rational
solution\footnote[3]{We thank Jacques-Arthur Weil for the remarkable
results (\ref{s1star}) and (\ref{N1}).}:
\begin{eqnarray}
\label{s1star}
S_{1}^{\ast }(w)\, = \,\,\,{\frac{{f(w)\cdot Q_{6}(w)}}{{w^{3}\cdot P_{7}(w)}}} 
\end{eqnarray}
where $f(w)=(1-w)(1+2w)(1-4w)^{5}(1+4w)^{3}(1+3w+4w^{2})$
and $\, Q_6$ is a polynomial of degree $\, 28$ given in ~\cite{ze-bo-ha-ma-04a}.

All these findings imply the following factorization of $L_{7}\,$ :
\begin{eqnarray}
\label{L1}
L_{7}\, &=&\,\,M_{6}\cdot L_{1}, \, \qquad \quad L_{7}\, =\,N_{6}\cdot N_{1} \\
\label{N1}
L_{7}\, &=&\,\,M_{1}\cdot L_{6},\qquad \quad L_{6}\,=\,L_{5}\cdot N_{1}
\end{eqnarray}
where $M_{6}$, $N_{6}$ and $L_{5}$ are operators of order respectively six, six
and five, and where the order-one operators $L_{1}$, $N_{1}$ and $M_{1}$ are such $%
L_{1}(S_{1})=0$, $N_{1}(S_{2})=0$ and $M_{1}^{\ast }(S_{1}^{\ast })=0$. The
operators $L_{6}$, $M_{6}$, $N_{6}$ and $L_{5}$ can easily be calculated
from the previous factorization relations, by right or left division by the
first order operators $L_{1}$, $N_{1}$ and $M_{1}$ defined above.
The decompositions (\ref{L1}, \ref{N1})
 imply that the general solution of (\ref{fuchs}) is a linear combination
 of $S_{1}$ and a solution, $S(L_{6})$,
of the sixth order linear homogeneous differential equation associated with $%
L_{6}$. The particular ``physical'' solution of $L_{7}$, $\tilde{\chi}^{(3)}$, can
be written as $\, 
\tilde{\chi}^{(3)}\,=\,\,\alpha \cdot S_{1}\,+\,S(L_{6}) $,  
where $\alpha $ is deduced from $L_{6}(\tilde{\chi}^{(3)}\,-\,\,\alpha \cdot
S_{1})=0$ and has the remarkably simple value $\alpha =1/3$. The homogeneous operator $%
L_{6}$ has the polynomial $w^{6}\,f(w)\,Q_{6}(w)$ in front 
of the highest derivative. 
The roots of the polynomial $Q_{6}(w)$ are also {\em apparent singularities} of the
linear homogeneous differential equation associated with $\, L_{6}$.

\section{Towards the new non apparent singularities: a Diff-Pad\'e method}
\label{diffPade}

From the ideas developped in Section 2, we were able to obtain a very
 long series expansion for $\tilde{\chi}^{(3)}$, sufficiently 
long to actually identify the Fuchsian equation for $\tilde{\chi}^{(3)}$.
However,
 when considering further
similar calculations on the next $\chi^{(n)}$'s ($\chi^{(4)}$,
 $\chi^{(5)}$, ...), we may not be able to obtain large enough
 series to identify the corresponding Fuchsian equations.
One may ask the following question: what kind of 
results can one get from the analysis of long series, but not long enough
to find the exact  Fuchsian equation.
This obviously supposes taking into account 
some of the ideas previously encountered, namely the homogeneous character
of the linear differential equation,
the Fuchsian character of some known singularities (like $\, s\, = \pm i$
 and Nickel's singularities, see (\ref{nick})),
the possible occurrence of unexpected,
 but simple, regular singularities (like the quadratic
 numbers $\, 1+\, 3\, w+ \, 4\, w^2\, = \, 0$), and 
above all, {\em the possible occurrence of 
quite a large set of apparent
 singularities}\footnote[4]{Which suggests considering
Fuchsian equations with quite large degrees for the polynomial coefficients,
compared to the order of the homogeneous linear differential equation.} (like
the 28 roots of polynomial $\, P_7$).

In fact the preliminary studies we performed before finding the Fuchsian
equation (\ref{fuchs}), give some hint as to the kind of results
one might expect~\footnote[5]{At this step one can certainely consider using
 the differential-Pad\'e approximation calculations 
developped by A.J. Guttmann {\it et al.}~\cite{diff2}, or similar differential-Pad\'e approximations
considered by other authors~\cite{diff}.}.

One knows that $\tilde{\chi}^{(3)}$ satisfies a linear differential equation. For a
given number of terms $N$ in the $\tilde{\chi}^{(3)}/w^9$ series, {\em there is} a
linear differential equation of order $q$ that reproduces the first $N-q$ terms but
may fail for subsequent coefficients.
Let us show what kind of information arises from the differential equations
 obtained for increasing order $q$, and increasing degrees of the polynomials
in front of the derivatives. We will consider the following form :
\begin{eqnarray}
\label{Xode}
\sum_{n=0}^q \, p_n(w) {\frac{d^n F(w)}{dw^n}}\, =\,\,0.
\end{eqnarray}
There are various ways to choose the degrees of the polynomials $p_n(w)$'s.
One may take:
\begin{eqnarray}
\label{degdeg}
deg(p_n\,)\,=\,deg(p_{n-1})+1, \quad \quad \quad deg(p_n)\,=\,\mu+n.
\end{eqnarray}
The differential equation (\ref{Xode}) then  has $1/2(q+1)(q+2\mu+2)$ unknowns.
The number $N$ of terms in the $\tilde{\chi}^{(3)}/w^9$ series used is:
\begin{eqnarray}
\label{hyperb}
N\, =\,\,  {\frac{1}{2}}(q+1)(q+2\mu+2) -1  -q.
\end{eqnarray}
One should note that the results below are not dependent on the form
(\ref{degdeg}) we took, and are also obtained when the degrees of the
polynomials $p_n(w)$ are taken to be
 equal\footnote[1]{The form (\ref{degdeg}) assures $w=\infty$ as a regular singularity
of the differential equation.}.

We systematically considered linear homogeneous differential equations
with polynomial coefficients (in $ w$) of increasing orders. 
The degrees of the various polynomial coefficients are taken as large
as possible. With such a strategy the order of the homogeneous 
linear differential equation, and 
the degrees of the polynomials, {\em are not on the same footing} (the order
 is much smaller than the degrees and the order of a differential 
equation is a much more fundamental character 
of the differential equation). For $N$ terms in the series, one then explores
all the linear differential equations reproducing these terms with $(q, \mu)$
staying on the hyperbola (\ref{hyperb}).

Let us fix $q=4$ and compute the linear differential equation for increasing values
of $\mu$. The polynomial in front of the highest derivative is solved for the
variable $w$. For $\mu=6$, one gets $w=-1/2$ with three correct digits, $w=-1/4$
(double) with five digits, $w=+1/4$ (double) with four digits, and $w=1$ with two
digits. With these values of $w$, we get four other solutions. As $\mu$ increases, the 
number of correct digits of the roots $w=-1/2$,
$w=-1/4$, $w=1/4$ and $w=1$ increases. For $\mu=15$, the values become correct
to, respectively, 8, 13, 14 and 6 digits. For $\mu=28$, they are correct to
14, 20, 26 and 13 digits. The roots $w=\pm 1/4$ still have multiplicity 2.
One observes that for $q=4$ and $\mu=28$, the number of series coefficients used
is just 150 terms. Further, as $\mu$ increases, the other solutions increase in number
($p_q(w)$ has $q+\mu$ roots) and do not converge to stable values. At 
around $\mu \sim 19$ a root oscillating around $-1/4$ appears. One can imagine
that the root $w=-1/4$ is going to be of multiplicity 3. For the root
$w=+1/4$, one has to wait until $\mu=28$ to see this happen.

The differential equations we obtain are quite cumbersome and carry as much
information (in the number of unknowns) as the $N$ terms used.
This is, then, just another way to {\it encode} the information contained in the $N$
terms. What is remarkable in this procedure is the appearance of the
singularities with increasingly correct digits and the fact that they converge to
the correct multiplicity as $\mu$ increases. What is more remarkable is
{\em the appearance of the other unexpected singularities}
$1+3w+4w^2\, =\, 0$ which were obtained with
three digits at $(q=4, \mu=2)$, and with nine correct digits at $(q=6, \mu=35)$.

Within this systematic approximation scheme,
we actually discovered, at smaller
orders than the actual one (namely order seven), the existence of the
 unexpected quadratic numbers singularities
 $\, 1+\, 3\, w+ \, 4\, w^2\, = \, 0$, 
with an extremely good level of confidence. This indicates that such a systematic 
procedure was probably converging towards some exact result, which we actually
 found with the next orders.
This method is obviously applicable to any series that satisfies a
linear homogeneous differential equation. Actually we will show, in a forthcoming
publication, that we can actually find the singularities of the successive
$\, \chi^{(n)}$'s even when we are not able to find the ODE's 
satisfied by the $\, \chi^{(n)}$'s.

\section{Conclusion}
\label{conclu}
Considering the isotropic Ising square lattice model susceptibility, we focused on the
third order component $\chi^{(3)}$ in the form given by Nickel \cite{nickel-99}.
We used an expansion in the variables $x_j$ that appear in the integrand of
the double integral defining $\chi^{(3)}$ instead of the variable $w$ (or $s$).
The angular integration
becomes straightforward thanks to the remarkable formula we derived for the
quantity $y_j\, x_j^n$. This formula is a Fourier series carrying one
summation index and where the coefficients are {\em non terminating} $_{4}F_{3} $
hypergeometric functions. We succeeded, in this way, to write $%
\tilde{\chi}^{(3)}$ as a fully integrated expansion, containing few sums of products
of three such hypergeometric functions. For $\tilde{\chi}^{(2)}$, the sums bear on
products of two hypergeometric functions but due to remarkable identities, 
they ``fuse'' into a {\em single hypergeometric function}
 giving the known result (see Appendix E and F). 
 We note that
these {\em fusion identities} on hypergeometric functions occurring for $%
\tilde{\chi}^{(2)}$, may be the sign of deep
structures and symmetries. We have not yet recognized
 a similar mechanism for $\tilde{\chi}^{(3)}$, but 
it seems plausible. 

We used the fully integrated multisum giving $\chi^{(3)}$ to generate a long
series of coefficients with a {\em polynomial time} ($N^4$) algorithm. We recall
that the
series coefficients are obtained {\em without any numerical approximation}. 
We were able to obtain the series up to order 490 in the variable 
$w$, and used it to find the seventh order linear 
differential equation (\ref{fuchs}) satisfied by $\chi^{(3)}$.
This limitation to 490 terms is a reflection of the minimal computer resources used. 
Much longer series can readily be obtained.
 This differential equation, which is of Fuchsian
type, is highly non trivial and structured. This result shows that 
quite involved sums of products of three hypergeometric functions can be solutions
of quite simple linear differential equations.

The ferromagnetic, antiferromagnetic, zero or infinite temperature points,
the non-physical and Nickel singularities are all regular singular points of
the differential equation. The Fuchsian differential equation shows other
simple singularities, which are not of Nickel's form, and are lying on the
circles $\vert s \vert=\sqrt{2}$ and $\vert s \vert={\frac{{1}}{\sqrt{2}}}$.
We also note the occurrence of a polynomial of degree 28 in $w$ whose roots
are {\em apparent singularities} of the Fuchsian differential equation.
We also saw
that any slight deformation of the Fuchsian differential equation
drastically modifies the growth of the coefficients in 
the series expansions of the solutions, confirming 
the highly selective character of this ordinary differential equation.

The behavior of the local solutions of the Fuchsian differential equation at
each regular singular point is sketched and shows the occurrence of
logarithmic terms up to $\log^2$. The question of whether these terms occur
in the physical solution $\tilde{\chi}^{(3)}$ is of the utmost importance and,
particularly, for the new singularities $1+3w+4w^2=0$ not previously 
known\footnote[2]{For $1+3w+4w^2=0$ singularities, in Section 3, we gave arguments
in favor of its absence in $\tilde{\chi}^{(3)}$.}.
We will report on the integration constants corresponding 
to the physical solution $\tilde{\chi}^{(3)}$ in a forthcoming publication.

The Fuchsian differential equation has two simple rational, and algebraic,
solutions enabling one to give the factorization of the differential operator
corresponding
to the Fuchsian differential equation. The solution is a linear
combination of a rational solution, namely $w/(1-4\,w)$, and 
of a solution of a sixth order homogeneous linear
differential equation.

Recall that the variable $w$ we used deals with both high and low
temperature cases on an equal footing. Furthermore, the key ingredient in obtaining
(\ref{khi3sergen}), and finally (\ref{fuchs}), is the Fourier expansion of
$y\,x^{n}$, a quantity appearing in any $\chi ^{(n)}$. So the expansion method
we used
is not specific to the third contribution $\chi^{(3)}$ and can be
generalized, mutatis mutandis, to the other $\chi^{(n)}$, and in
particular
 the ones with $\, n$ even, associated with the {\em low-temperature} 
susceptibility (in Appendix E we discuss the form of $\chi ^{(2)}$
in this variable). The next step is clearly the evaluation of $\chi ^{(4)}$
which can be seen \cite{nickel-99,nickel-00} to be
a function of $s^2$, and, in fact, a function of $\, w^2\, $.
This gives some hope of finding the corresponding Fuchsian differential
equation.

The Fuchsian differential equation, satisfied by $\chi^{(3)}$,
is certainly an important step towards the understanding of the 
analytical structure
of isotropic Ising square lattice model susceptibility. Along these lines, one can easily
imagine  finding the Fuchsian differential equation satisfied by $\chi^{(3)}$
for the {\em isotropic triangular} Ising model.
However, recalling
the nice and simple exact results we displayed in the introduction for 
the {\em anisotropic} square Ising model, on the 
anisotropic triangular Ising model, and on the   
checkerboard Ising model, it is tempting to seek 
for the homogeneous ``Fuchsian partial differential equation'' (generalizing our Fuchsian 
differential equation)
in the corresponding two, three or four, high temperature variables  $t_1$, $t_2$,
$t_3$ and $t_4$. No doubt the modulus $k$ of the
 elliptic functions parameterizing the Ising model
 (see (\ref{modulus})) should play a special role
in writing down a simple enough expression for this Fuchsian PDE. 
Having such a Fuchsian PDE, it will be extremely interesting to
see how this homogeneous Fuchsian PDE can be invariant\footnote{See for instance~\cite{Sasa}.}
 (or covariant since it is homogeneous) under the
inversion relation symmetries (\ref{kiinv}),(\ref{inverse}) and the
 whole set of {\em birational transformations}~\cite{birat,birat2,birat3} 
generated by these inversion relation symmetries
combined with the geometrical symmetries.

\hskip 2cm

\textbf{Acknowledgments} We thank Jacques-Arthur Weil for valuable
comments and calculations on our Fuchsian equation. J-M.M would like to
thank A. J. Guttmann and W. Orrick for extensive discussions
 and for a large number of e-mail exchanges and calculations more recently. We
would like to particularly  thank A. J. Guttmann for careful checks on our
series. (S. B) and (S. H) acknowledge partial support from PNR3-Algeria.

\section{Appendix A}
\label{appena}
In this Appendix, we give the various formulas used to obtain
(\ref{khi3sergen}). We use $<\cdots >$ to mean the 
normalized angular integrations
 $\, \prod_i {{d\phi_i} \over {2\, \pi}}$.
The variables $x$ and $y$ are defined as in (\ref{varxy2}). From (\ref%
{FourExp}), one gets: 
\begin{eqnarray}
<yx^{n}\,\cos (p\phi )>\,\,\,=\,\,w^{p}\,a(n,p)\,\,=\,\,w^{p}\,a(p,n).
\end{eqnarray}
Recall that $a(n,p)$ is defined in (\ref{defa}) in terms of $%
_{4}F_{3} $ hypergeometric functions.

Define the following functions: 
\begin{eqnarray}
g(n,p) &=& \sum_{i=0}^\infty (i+1)\cdot   w^i a(n+i+1,p) =
w^{-p} \,  < {\frac{y\,x^{n+1}}{(1-wx)^2}} \, \cos(p\phi) >, \nonumber \\
v(n,p) &=& \sum_{i=0}^\infty w^i  a(n+i+1,p) =
w^{-p} \, < {\frac{y\,x^{n+1}}{1-wx}} \, \cos(p\phi)>. 
\end{eqnarray}
From the integral representations, one has:
\begin{eqnarray}
a(n+1,p) &=& v(n,p) -w\, v(n+1,p),  \nonumber \\
v(n,p) &=& g(n,p) -w\, g(n+1,p). \nonumber 
\end{eqnarray}

The following integrals can, then, be computed 
\begin{eqnarray}
<yx^{n}\,\cot (\phi /2)\,\sin (k\phi )> &=&\,y_{0}x_{0}^{n}\,-w^{k}\cdot
\Bigl(a(k,n)+2wv(k,n)\Bigr), \nonumber \\
<yx^{n}\,\cot (\phi /2)^{2}\,\cos (k\phi )>
&=&-2 k\, y_{0}x_{0}^{n}-w^{k} \, \Bigl(a(k,n)+4w g(k,n)\Bigr) 
\nonumber \\
&&\quad +y_{0}\,x_{0}^{n} \, <{\frac{1}{\sin (\phi /2)^{2}}}>.  \label{singg}
\end{eqnarray}
where $x_{0}$ and $y_{0}$ are the variables defined in (\ref{varxy2}) and
taken at $\phi =0$.

Some integrals appearing in the intermediate steps are ($k\geq 1$):
\begin{eqnarray}
<{\frac{1-\cos (k\phi )}{\sin (\phi /2)^{2}}}\cos (p\phi )> &=&\left\{ 
\begin{array}{ll}
2(k-p), &\quad k>p \\ 
0 &\quad  k\leq p
\end{array}
\right. \\
<\cot (\phi /2)\sin (k\phi )\cos (p\phi )> &=&\left\{ 
\begin{array}{ll}
1, &\quad  k>p \\ 
1/2, &\quad k=p \\ 
0 & \quad k<p
\end{array}
\right.
\end{eqnarray}

We now turn to the ``artificial'' singularity appearing on the right hand side of (%
\ref{singg}). This term contributes to the total integrand (\ref{chi3tild})
as:
\begin{eqnarray}
<{\frac{1}{\sin (\phi /2)^{2}}}>\,\Bigl(d(0)+2\sum_{p=1}^{\infty
}w^{3p}\,d(p)\Bigr) \qquad \qquad \mbox{with}  \label{singpart}
\end{eqnarray}%
\begin{eqnarray}
d(p) &=&y_{0}x_{0}^{p}\,  \Bigl(S(p+2,p)-S(p+1,p+1)\Bigr), \nonumber  \\
\label{Spq}
S(p,q) &=&a(p,0)a(q,0)\,+2\sum_{k=1}^{\infty }w^{2k}\,a(p,k)a(q,k).
\end{eqnarray}
Noting that 
\begin{eqnarray}
\label{Spqyx}
S(p,q)\,=\,\,<y^{2}x^{p+q}>\,\,=\,\,<yx^{p}\,yx^{q}>  
\end{eqnarray}%
it is straightforward to see the cancellation of $d(p)$ and, thus, of (\ref%
{singpart}) as it should.

Note that formula (\ref{FourExp}) has been used extensively in the
following appendices, and each time we have an expression similar to (\ref{Spqyx}),
we manage to write it in terms of (\ref{FourExp}). Another trick used throughout
the appendices is the technique of switching 
to-and-fro between the sums and the integral representation,
and the judicious selection of the correct place to use (\ref{yx}).

\section{Appendix B}
\label{appenb}
Expansion of (\ref{chi3tild}) in terms of $y_{j}x_{j}^{n}$,
using (\ref{FourExp}) and the integration rules given in Appendix A, gives
for $\tilde{\chi}^{(3)}$: 
\begin{eqnarray}
\tilde{\chi}^{(3)}\,=\,\,\, 8\ w^{5}\,  \Bigl(I(0)+2\sum_{p=1}^{\infty }w^{3p}I(p)%
\Bigr),  \label{khi3form1}
\end{eqnarray}
with
\begin{eqnarray}
I(p) &=&{\frac{1}{w^{2}}} \, <y_{1}y_{2}y_{3}\,(x_{1}x_{2}x_{3})^{p}\,H^{(3)}>
\nonumber \\
&=&\, C(p)+w^{2}F(0,p)+2w^{2}\sum_{k=1}^{\infty }w^{3k}F(k,p),
\end{eqnarray}%
where $H^{(3)}$ is taken in the form (\ref{h3new}).
$F(k,p)$ consists of sums of cubic terms of the hypergeometric function 
(\ref{defa}), and can be written in terms of this hypergeometric
function and the auxillary functions $g(k,p)$ and $v(k,p)$ defined in
Appendix A, as:
\begin{eqnarray}
\label{Fkp}
F(k,p) &=&g(k+1,p)\, \Bigl(a(k,p+1)^{2}-a(k,p+2)a(k,p)\Bigr)  \nonumber \\
&& + a(k,p+2)\, v(k,p)^{2}\, +a(k,p)\, v(k,p+1)^{2} \nonumber \\
&& -2\, a(k,p+1)\, v(k,p)\, v(k,p+1).  
\end{eqnarray}
The function $C(p)$, written as:
\begin{eqnarray}
\label{defC}
C(p)\, =\, -\frac{1}{2}\, y_{0}^{2}x_{0}^{2p}\, C_1(p)\, +y_{0}x_{0}^{p}\, C_2(p),  
\end{eqnarray}
consists of a linear term  in the $a$'s:
\begin{eqnarray}
\label{defC1}
C_1(p)\,=\,\, a(0,p+2)+x_{0}^{2}a(0,p)-2x_{0}a(0,p+1),  \nonumber 
\end{eqnarray}
and a quadratic term in the $a$'s:
\begin{eqnarray}
\label{defC2}
C_2(p)\,=\,\, \sigma _{0}(p)+2w\,\sigma _{1}(p)\, +2 w\, x_{0}\,\sigma _{2}(p),
\end{eqnarray}
where the sums $\sigma _{i}(p)$ $(i=0, 1, 2)$ are given by:
\begin{eqnarray}
\sigma _{0}(p) &=&\sum_{k=1}^{\infty }\,(k+1)\,w^{2k}\, \Bigl(%
a(k,p+1)^{2}\,\,-a(k,p+2)\,a(k,p)\Bigr), \nonumber  \\
\sigma _{1}(p) &=&\sum_{k=1}^{\infty }w^{2k}\, \Bigl(a(k,p+1)%
\,v(k,p+1)-a(k,p+2)\,v(k,p)\Bigr), \nonumber  \\
\sigma _{2}(p) &=&\sum_{k=1}^{\infty }w^{2k}\, \Bigl(a(k,p+1)\,v(k,p)%
\,-a(k,p)\,v(k,p+1)\Bigr). \nonumber 
\end{eqnarray}

To get the final form of $\tilde{\chi}^{(3)}$, further manipulations are
made on that part of $I(0)$ and $I(p)$ in (\ref{khi3form1}) containing the
$C $'s. The aim is to get {\em only one summation} instead of two.
We show in Appendix C how to obtain the part of (\ref{khi3form1}) containing
the $C$'s. It reads:
\begin{eqnarray}
&&C(0)\, +2\sum_{p=1}^{\infty }w^{3p}C(p)\,=
\,\,\, \xi_{1}\,+\sigma _{C} \qquad \quad  \mbox{where } \label{jvd} \\
&&\xi_{1}\,=\,\,y_{0}\, C(0)  
+\frac{y_{0}^{3}}{w}\cdot
\Bigl( J_0(0)-a(1,0) \Bigr)  \label{closed1}
\end{eqnarray}
\begin{eqnarray}
\label{formsigmaC}
\sigma_{C}\,=\,\,\,2\,y_{0}^{3}\, \sum_{n=1}^{\infty } w^{3n-2}x_{0}^{n-1}\,
\Bigl( J_{0}(n)\,-w \, (1-wx_0)\cdot J_1(n) \Bigr) \nonumber 
\end{eqnarray}
\begin{eqnarray}
\label{J0}
\mbox{where} \qquad
 J_{0}(n)\,\, =\,\, a(0,n)^{2}-w^{2}a(1,n)^{2},  \nonumber 
\end{eqnarray}%
\begin{eqnarray}
\label{J1}
J_{1}(n)\,=\,\, (a(0,n)+wa(1,n))\,(a(0,n+1)-x_{0}a(0,n)).  
\end{eqnarray}

We have decomposed the term $\sigma _{C}$ as:
\begin{eqnarray}
\sigma _{C}\,=\,\, \xi_{2}+2w^{8}\sigma   \qquad  \qquad  \qquad
 \mbox{with} \nonumber
\end{eqnarray}
\begin{eqnarray}
\label{closed2}
&&\xi_{2}\,=\,2y_{0}^{3}\cdot \Bigl(w \cdot
( J_{0}(1)\, -w \cdot(1-wx_0)\cdot J_1(1))  \nonumber \\
&&\quad \quad
+x_{0}w^{4}\cdot ( J_{0}(2)\, -w\cdot(1-wx_0)\cdot J_1(2))
\,+x_{0}^{2}w^{7}\,a(0,3)^2 \Bigr), \nonumber 
\end{eqnarray}
and
\begin{eqnarray}
\label{defsigma}
\sigma \,&=&\,\,y_{0}^{3}\cdot \sum_{p=0}^{\infty }\,w^{3p}\,x_0^{p+2}\cdot \\
&& 
\Bigl(
(wx_0-1)\,J_{1}(p+3)\, -w\,a(1,p+3)^2\, +w^2\, x_0\,a(0,p+4)^2
\Bigr). \nonumber 
\end{eqnarray}

Collecting all previous results, and after some manipulations,
$\tilde{\chi}^{(3)}$ can then be written as: 
\begin{eqnarray}
\tilde{\chi}^{(3)}\,=\,\,\,8\,w^{9}\cdot \Bigl(\tilde{\chi}_{cl}^{(3)}+\,2w^{4}%
\cdot \Xi \Bigr), \nonumber 
\end{eqnarray}
where 
\begin{eqnarray}
\label{defkhi3closed}
\tilde{\chi}_{cl}^{(3)}\,=\,\,\frac{1}{w^{4}}(\xi_{1}+\xi_{2})+\frac{F(0,0)}{%
w^{2}}+2w\cdot (F(1,0)+F(0,1)),  
\end{eqnarray}
and
\begin{eqnarray}
\label{Xi}
\Xi\, &=& \,\,\sigma +\sum_{p=0}^{\infty }w^{3p}\cdot \Bigl( F(0,p+2)+F(p+2,0) \Bigr) \nonumber \\
&&\quad +2\sum_{k=0}^{\infty }\sum_{p=0}^{\infty}w^{3k+3p}\cdot F(k+1,p+1). 
\end{eqnarray}
Appendix D gives the detailed derivation of the closed form
(\ref{khi3closed}) from (\ref{defkhi3closed}).

\section{Appendix C}
\label{appenc}
The purpose in this Appendix is to derive relation (\ref{jvd}).
From the definition of the variables $x_{n}$ and $Z_{n}=\exp (i\phi _{n})$,
one can deduce the following identity:
\begin{eqnarray}
\label{key}
(Z_{1}-Z_{2})\, \frac{w^{2}x_{1}x_{2}}{1-w^{2}x_{1}x_{2}}
\, =\,\,  w \cdot (x_{2}-x_{1})\, \frac{%
Z_{1}Z_{2}}{1-Z_{1}Z_{2}}.   
\end{eqnarray}

The first step is to simplify the expression of $C_2(p)$ defined in
(\ref{defC2}). Taking the integral representation of $a(k,n)$ and $v(k,n)$
($n=p, \, p+1, \, p+2$), and summing on the index $k$, the sums $\sigma _{i}(p)$
become:
\begin{eqnarray}
\sigma _{0}(p) = w^{-2p}<y_{1}y_{2}\frac{x_{1}x_{2}}{1-w^{2}x_{1}x_{2}}%
\frac{2-w^{2}x_{1}x_{2}}{1-w^{2}x_{1}x_{2}}%
(Z_{1}Z_{2})^{p}(Z_{1}Z_{2}-Z_{1}^{2})>, \nonumber 
\end{eqnarray}
\begin{eqnarray}
\sigma _{1}(p) = w^{-2p}<y_{1}y_{2}\frac{x_{1}x_{2}}{1-w^{2}x_{1}x_{2}%
}\frac{x_{2}}{1-wx_{2}}(Z_{1}Z_{2})^{p}(Z_{1}Z_{2}-Z_{1}^{2})>, \nonumber 
\end{eqnarray}
\begin{eqnarray}
\sigma _{2}(p) = {{1}\over{2}} w^{-2p}<y_{1}y_{2}\frac{x_{1}x_{2}}{%
1-w^{2}x_{1}x_{2}}\frac{1+wx_{2}}{1-wx_{2}}%
(Z_{1}Z_{2})^{p}(Z_{1}-Z_{2})>. \nonumber 
\end{eqnarray}

The sum $C_2(p)$ (sum of the $\sigma$'s) becomes:
\begin{eqnarray}
\label{C2pZ}
&& C_{2}(p) = w^{-2p} \cdot \Bigl(  
<y_{1}y_{2}\frac{x_{1}x_{2}}{(1-w^{2}x_{1}x_{2})^{2}}(Z_{1}Z_{2})^{p}(-%
\frac{1}{2}(Z_{2}-Z_{1})^{2})>+ \nonumber \\ 
&& <y_{1}y_{2}\frac{1+wx_{2}}{1-wx_{2}}\frac{x_{1}x_{2}}{%
(1-w^{2}x_{1}x_{2})}(Z_{1}Z_{2})^{p}(Z_{1}-wx_{0})(Z_{2}-Z_{1})> 
\Bigr), 
\end{eqnarray}
where, for the first term in the brackets at the right hand side of
(\ref{C2pZ}), the quantity $Z_{1}Z_{2}-Z_{1}^{2}$ has been replaced by
$-\frac{1}{2}(Z_{2}-Z_{1})^{2}$ due to the symmetry of the rest of the
integrand in the integration variables.
Using (\ref{key}) and the following relation:
\begin{eqnarray}
<\frac{y}{x}\, Z^{n+1}>\,\, =\,\,\, w^{n+3}a(1,n+1), \qquad \qquad  n\geq 0 \nonumber 
\end{eqnarray}
then expanding in the\ $Z$ variables, working out the integration over
the angles and making use of the identities:
\begin{eqnarray}
\label{def0p}
<y\, \frac{1+wx}{1-wx}\cdot \cos (p\phi )>\,\,\, =\,\, \, <y\, \frac{1+wx}{1-wx}%
Z^{p}>\,\, =\,\, \, y_{0}x_{0}^{p}w^{p},  
\end{eqnarray}
\begin{eqnarray}
\label{defv0p}
v(0,p)\, =\, \frac{1}{2w} \cdot (y_{0}x_{0}^{p}-a(0,p)),  
\end{eqnarray}%
one obtains
\begin{eqnarray}
\label{C2simple}
C_{2}(p)\, =\,\,  S_{0}(p)+S_{1}(p)+
\sum\limits_{j=1}\,w^{2j-1}\,y_{0}x_{0}^{j+p}\, J_{2}(j+p), 
\end{eqnarray}
with
\begin{eqnarray}
\label{S0}
S_{0}(p) = \sum\limits_{j=1}\,j\,w^{2j}\,J_{0}(j+p+1), \qquad   
S_{1}(p) = \sum\limits_{j=1}\,w^{2j-1}\,J_{1}(j+p).  \nonumber 
\end{eqnarray}
$J_0$ and $J_1$ are defined in (\ref{J1}), while $J_2$ reads:
\begin{eqnarray}
\label{j2}
J_{2}(n) &=&x_{0} \Bigl( a(0,n)-wa(1,n) \Bigr)
-\Bigl( a(0,n+1)-wa(1,n+1) \Bigr). \nonumber 
\end{eqnarray}
$C_2(p)$ is now written as a sum of contributions quadratic in $a$'s and linear
in $a$'s (the last term). The latter will be shown to cancel $C_1(p)$  identically
in (\ref{defC}). Using the following relation:
\begin{eqnarray}
\label{A1n}
2w^{2}a(1,n)\, =\,\, \,  a(0,n)\, -a(0,n-1)\, -w^{2}a(0,n+1)  
\end{eqnarray}
one obtains the identity:
\begin{eqnarray}
\label{J2eq2}
J_{2}(n)\, =\, \, {\frac{1}{2wx_{0}}} \Bigl( C_1(n-1)-w^{2}x_{0}C_1(n) \Bigr).\nonumber 
\end{eqnarray}
The last sum in (\ref{C2simple}) simply reads:
\begin{eqnarray}
\label{S2eq2}
\sum\limits_{j=1}\,w^{2j-1}\,y_{0}x_{0}^{j+p}\cdot J_{2}(j+p)\, = \, \, 
\frac{1}{2} y_{0} x_{0}^{p}\, C_1(p).  \nonumber 
\end{eqnarray}
Collecting terms in (\ref{defC}), one gets:
\begin{eqnarray}
\label{Csimple}
C(p)\, =\,\,  y_{0}x_{0}^{p} \cdot \Bigl( S_{0}(p)+S_{1}(p) \Bigr)  
\end{eqnarray}

Some manipulations, such as shifts in the indices of the sums, allow us to evaluate 
two summations and to cast the quantity of interest (\ref{jvd}) as:
\begin{eqnarray}
\label{RC}
C(0)\, +2\sum\limits_{p=1}^{\infty }\,w^{3p}\,C(p)\, =\,\, \,
y_{0}\,C(0)\,-2\,w\,y_{0}^{3}\cdot \xi_{0}\,+\sigma _{C}.
\end{eqnarray}

$\xi_0$ is a sum that can be worked out using 
(\ref{yx}), (\ref{FourExp}), (\ref{Spq}) and (\ref{Spqyx}):
\begin{eqnarray}
\label{Cl0}
\xi_{0} = \sum\limits_{n=0}^{\infty }\,w^{2n}\,J_{0}(n+1)\,=\,\,
 {\frac{1}{2w^2}} \Bigl( a(1,0)-J_0(0) \Bigr). 
\end{eqnarray}
This ends the derivation of (\ref{jvd}) given in Appendix B.

\section{Appendix D}
\label{append}
In the closed part (\ref{defkhi3closed}), there is still a
summation to be done in the term $\xi_{1}$ defined in (\ref{closed1}),
namely the evaluation of $C(0)$. For this purpose, we consider
$C(p)$ in the form (\ref{Csimple}).

To evaluate $S_{0}(0)$, one shifts the index of summation to cast it as:
\begin{eqnarray}
S_{0}(0)\, =\,\, \sum\limits_{n=1}\,(n-1)\,w^{2n-2}\,J_0(n)\,
=\,\,\, \widehat{S_{0}}(0)\,-\xi_{0},
\end{eqnarray}
with
\begin{eqnarray}
\widehat{S_{0}}(0)\, =\, \, \,
\sum\limits_{n=1}^{\infty }\,n\,w^{2n-2} \,J_0(n). \nonumber 
\end{eqnarray}

At this point,  note that $x^{n}$ has the following Fourier expansion:
\begin{eqnarray}
x^{n}\,=\,\,b(n,0)\,+2\sum_{k=1}^{\infty }w^{k}b(n,k)\,\cos {k\phi }, \nonumber 
\end{eqnarray}%
where $b(k,n)$ reads (with $m=k+n$): 
\begin{eqnarray}
\label{defbkn} 
&& b(k,n) =\, \, {m-1 \choose k} \cdot \\
&& {_{4}}F_{3}\Bigl({\frac{{(1+m)}}{{2}}},{%
\frac{{(1+m)}}{{2}}},{\frac{{(2+m)}}{{2}}},{\frac{{m}}{{2}}}; 
 1+k,1+n,1+m;16w^{2}\Bigr)  \nonumber 
\end{eqnarray}
and satisfies:
\begin{eqnarray}
\label{b1n}
b(i,p) &=& \, a(i-1,p)\, -w^{2}a(i+1,p),\, \qquad \qquad  i \geq 1  \\
n\,  b(1,n) &=&\, b(n,1).  \nonumber
\end{eqnarray}
Using (\ref{b1n}) and the symmetry of $a(k,p)$ in $k$ and $p$,
$\, \widehat{S_{0}}(0)$ becomes:
\begin{eqnarray}
\widehat{S_{0}}(0)\, =\,\,  \frac{1}{2}\sum\limits_{n=1}^{\infty }\,w^{2n-2} \cdot 
\Bigl(  a(n,0)b(n,1)\,-2nw^{2}a(n,1)^{2}\,  \nonumber \\ 
    \qquad   \qquad      +  n\, a(n,0) \cdot (a(n,0)+w^{2}a(n,2) \Bigr), \nonumber 
\end{eqnarray}
and, in integral form, we obtain:
\begin{eqnarray}
&&\widehat{S_{0}}(0) = \frac{1}{2w^{2}}
< y_{1}\frac{wx_{1}x_{2}}{1-w^{2}x_{1}x_{2}}\cos (\phi _{2})  > \nonumber  \\ 
&& \quad \quad + \frac{1}{2 w^{2}}  
<  y_{1}y_{2}\frac{w^{2}x_{1}x_{2}}{(1-w^{2}x_{1}x_{2})^{2}}2\cos (\phi
_{2})\left( \cos (\phi _{2})-\cos (\phi _{1})\right) >. \nonumber 
\end{eqnarray}
Using the identity (\ref{key}) and relation (\ref{yx}),
$\widehat{S_{0}}(0)$ simply becomes:
\begin{eqnarray}
\label{S0tilda}
\widehat{S_{0}}(0)\,=\,\, \frac{1}{2w}\left\langle y_{1}y_{2}\,x_{2}\cos (\phi
_{2})\right\rangle\, =\,\,\frac{1}{2}\,a(0,0)\,a(1,1).  
 \nonumber 
\end{eqnarray}
For the evaluation of $S_{1}(p)$, we make use of (\ref{A1n}) to get:
\begin{eqnarray}
 S_{1}(0) &=& -{{1}\over{2}} \sum_{n=1} w^{2n-2} \cdot \Bigl(
 a(0,n+1) \cdot ( a(0,n-1)+w^2 a(0,n+1)) \nonumber \\
&& + a(0,n) \cdot  ( 2x_0w (a(0,n)+w a(1,n)) -(1+2w)a(0,n+1))
 \Bigr). \nonumber 
\end{eqnarray}
 Using (\ref{FourExp}), (\ref{Spq}), (\ref{Spqyx}) and the useful identities:
\begin{eqnarray}
\label{cos1}
<y^{2}\cos (\phi )>\,=\,\, \, 2 w\, a(0,0)a(0,1)\, +2\sum\limits_{n=1}^{\infty }\,%
w^{2n+1}\, a(0,n)\, a(0,n+1),   \nonumber 
\end{eqnarray}
\begin{eqnarray}
\label{cos2}
<y^{2}\cos (2\phi )>\, =\,\, \, w^{2}a(0,1)^{2}\, +2\sum\limits_{n=1}^{\infty }\,%
w^{2n}\, a(0,n-1)\, a(0,n+1),   \nonumber 
\end{eqnarray}%
we obtain
\begin{eqnarray}
S_{1}(0) &=&\frac{x_{0}}{2w} \, ( (a(0,0)+wa(1,0))\,%
\cdot a(0,0)\, -<y^{2}\, (1+wx)>)  \nonumber \\
&&+\frac{1}{4w^{3}}<y^{2}\cos (\phi )(1-2w\cos (\phi )+2w)>  \nonumber \\
&&+\frac{1}{4w^{2}}(3w^{2}a(0,1)^{2}\, 
+a(0,0)^{2}\, -2\, (1+2w)\, a(0,0)\, a(0,1)). \nonumber 
\end{eqnarray}%
With the relation (\ref{yx}) and using (\ref{defv0p}), we obtain:
\begin{eqnarray}
<y^{2}\cdot (1 +w x)>\, \,  =\,\, \,
v(0,0)\,=\,\, \,\frac{y_{0}\, -a(0,0)}{2w} .\nonumber
\end{eqnarray}%
With the following relation:
\begin{eqnarray}
1 -2\, w\,\cos (\phi )\,=\,\,\frac{1}{x}\, +w^{2} x, \nonumber
\end{eqnarray}%
and using (\ref{def0p}), we get:%
\begin{eqnarray}
<y^{2}\, \cos (\phi )\, (1-2w\cos (\phi )+2w)> \,=\, w y_{0}x_{0} \nonumber
\end{eqnarray}
and finally:%
\begin{eqnarray}
S_{1}(0) &=& \,\frac{1}{4w^{2}} \Bigl( 
x_{0}a(0,0)+(1 +2wx_{0})\, a(0,0)^{2}  \\
&& \qquad +3 w^{2}a(0,1)^{2} +2(w^{2}x_{0} -1-2w)\cdot a(0,0)a(0,1)
\Bigr). \nonumber 
\end{eqnarray}

The last step in calculating the closed form (\ref{defkhi3closed}) is to
explicitly evaluate the terms $F(0,0)$, $F(1,0)$ and $F(0,1)$. This can be done
with the help of Appendix A. Note that $g(1,1)$, and $g(0,0)$, can be written as:
\begin{eqnarray}
g(1,1) &=& \,{\frac{1}{2w^2}} \Bigl( g(1,0)-2g(0,0)+2v(0,0)-a(1,0) \Bigr), \nonumber \\
g(0,0) &=&\, \frac{1}{4w}\,\left( \frac{\tilde{E}\,\,}{1-4w}-\tilde{K}%
\,\right), \nonumber
\end{eqnarray}
$\tilde{E}$ and $\tilde{K}$ being defined in (\ref{KE}).

Collecting all the previous results, expanding in the basis of $\tilde{E}$ and $%
\tilde{K}$ functions, one obtains the relation (\ref{khi3closed}), where the
polynomials $q_{i,j}$ and $p_{i,j}$ are given by: 
\begin{eqnarray}
q_{00} &=&-96{w}^{6}+276{w}^{5}-120{w}^{4}+144{w}^{3}-64{w}%
^{2}+32w-16,  \nonumber \\
q_{10} &=&128{w}^{7}-832{w}^{6}-792{w}^{5}+740{w}^{4}-680{w}%
^{3}+295{w}^{2} \nonumber \\
&& +32w-16,  \nonumber \\
q_{01} &=&68{w}^{5}-212{w}^{4}-120{w}^{3}+89{w}^{2}-96w+48, 
\nonumber \\
q_{20} &=&960{w}^{7}-416{w}^{6}+1924{w}^{5}-1048{w}^{4}-170{w}%
^{3}+98{w}^{2} \nonumber \\
&& +3w-2,  \nonumber \\
q_{11} &=&584{w}^{5}-256{w}^{4}+744{w}^{3}-386{w}^{2}-38w+20, 
\nonumber \\
q_{02} &=&82{w}^{3}-32{w}^{2}+67w-34,  \nonumber \\
q_{30} &=&(4w-1)^{2}(4w+1)^{2}(2w-1)(2w+1) w^{2},  \nonumber \\
q_{21} &=&-(4w-1)(4w+1)(4w^{3}-28w^{2}-2w+5) w^{2} , \nonumber \\
q_{12} &=&-8\, (6w^{3}-9w^{2}-w+1) w^{2},  \nonumber \\
q_{03} &=&-2\, (3w-2) w^{2},  \nonumber \\
&&  \nonumber \\
p_{00} &=&528{w}^{6}-464{w}^{5}+248{w}^{4}-272{w}^{3}+96{w}%
^{2}-64w+16,  \nonumber \\
p_{10} &=&-1280{w}^{7}+352{w}^{6}+1984{w}^{5}-1504{w}^{4}+1284{w}%
^{3}-264{w}^{2} \nonumber \\
&& -64w+16,  \nonumber \\
p_{01} &=&-320{w}^{5}+208{w}^{4}+284{w}^{3}-184{w}^{2}+192w-48, 
\nonumber \\
p_{20} &=&256{w}^{8}-1280{w}^{7}+2320{w}^{6}-4144{w}^{5}+928{w}%
^{4}+372{w}^{3} \nonumber \\
&& -98{w}^{2}-7w+2,  \nonumber \\
p_{11} &=&384{w}^{6}-896{w}^{5}+960{w}^{4}-1548{w}^{3}+356{w}%
^{2}+78w-20,  \nonumber \\
p_{02} &=&56{w}^{4}-120{w}^{3}+94{w}^{2}-135w+34.  \nonumber
\end{eqnarray}

\section{Appendix E}
\label{appenf}
This Appendix contains the details of evaluation of $\chi^{(2)}$ using
the $yx^{n}$ expansion method. One should note however, that
$\chi ^{(2)}$ can be obtained in a straightforward manner by direct
integration.

One writes $\chi ^{(2)}$ as \cite{nickel-00}: 
\begin{eqnarray}
\chi^{(2)}\,=\,\,\Bigl(1-s^{-4}\Bigr)^{1/4}\,  \tilde{\chi}^{(2)} 
\end{eqnarray}
where 
\begin{eqnarray}
\tilde{\chi}^{(2)}\,=\,\,4 w^{4} \cdot < y^{2}x^{2}{\frac{1+w^{2}x^{2}}{(1-w^{2}x^{2})^3}}%
\cdot \Bigl(1-\cos (2\phi )\Bigr)\, >. \nonumber 
\end{eqnarray}
One defines:
\begin{eqnarray}
\label{grandA}
A(k,n)\, =\,\,A(-k,n)\,=\,\, w^{\left\vert k\right\vert }\,  a(\left\vert
k\right\vert,n ) \nonumber 
\end{eqnarray}
With the variable $Z=\exp (i\phi )$, the expansion of $yx^{n}$ is written as: 
\begin{eqnarray}
yx^{n}\,=\,\sum_{k=-\infty }^{\infty }A(k,n)\,Z^{k}\,=\,\,\sum_{k=-\infty
}^{\infty }A(k,n)\,Z^{-k} \nonumber 
\end{eqnarray}
Expanding in $x$, $\tilde{\chi}^{(2)}$ becomes :
\begin{eqnarray}
\tilde{\chi}^{(2)}\,=\,\,4 w^{4}\, \sum_{n=0}^{\infty }(n+1)^{2}\,
w^{2n}\, <y^{2} x^{2n+2} (1-\cos (2\phi ))>.  \label{khi2expansion}
\end{eqnarray}

Now write $<y^{2}x^{2n+2}\,(1-\cos (2\phi ))>$ as $<yx^{n_{1}}
\,yx^{n_{2}}\,(1-\cos (2\phi ))>$ with $n_{1}+n_{2}=2n+2$. One has: 
\begin{eqnarray}
&<&yx^{n_{1}}yx^{n_{2}}(1-\cos (2\phi ))>  \label{eq2} \\
&=&\sum_{k_{1},k_2=-\infty }^{\infty }
A(k_{1}, n_{1})A(k_{2},n_{2})<(1-Z^{2}/2-Z^{-2}/2)\,Z^{k_{1}+k_{2}}>
\nonumber \\
&=&\sum_{k=-\infty }^{\infty }A(k, n_{1})\Bigl(%
A(k, n_{2})-A(k+2, n_{2})/2-A(k-2, n_{2})/2\Bigr).  \nonumber
\end{eqnarray}%
Some manipulations give: 
\begin{eqnarray}
\label{lastforchi2}
&<&yx^{n_{1}}\,yx^{n_{2}}\cdot(1-\cos (2\phi ))>  \label{eq6} \\
&=&\sum_{k=1}^{\infty }\Bigl(A(k-1, n_{1})-A(k+1, n_{1})\Bigr)\Bigl(%
A(k-1, n_{2})-A(k+1, n_{2})\Bigr).  \nonumber
\end{eqnarray}
Coming back to the definition (\ref{grandA}) of $A(k,n)$, one recognizes
in (\ref{lastforchi2}) the function $b(k,n)$ defined in (\ref{defbkn}). 
$\tilde{\chi}^{(2)}$, then, becomes (recall that $n_1+n_2=2n+2$) 
\begin{eqnarray}
\tilde{\chi}^{(2)}\, =\, \, 4w^4 \cdot \sum_{n=0}^\infty \sum_{k=0}^\infty
(n+1)^2 w^{2n+2k} \, b(k+1,n_1)\,  b(k+1,n_2). \nonumber 
\end{eqnarray}

One thus has many equivalent forms for $\tilde{\chi}^{(2)}$ depending on the
partition $n_1+n_2=2n+2$. The above sums merges into a {\em single} hypergeometric
function:
\begin{eqnarray}
\label{chi2F}
\tilde{\chi}^{(2)}\,=\,\,4\,w^{4}\cdot {_{2}}F_{1}\Bigl(
{{5}\over{2}},{{3}\over{2}};3;16w^{2}\Bigr),  
\end{eqnarray}
which is the well-known result \cite{wu-mc-tr-ba-76}:
\begin{eqnarray}
\label{khi2}
\tilde{\chi}^{(2)}\,=\,\,{\frac{{1}}{{6\pi }}} \cdot 
\Bigl({\frac{1-8w^{2}}{%
1-16w^{2}}}\cdot E(4w)\, -K(4w)\Bigr).  
\end{eqnarray}

\section{Appendix F}
\label{appeng}
Here we give an alternative derivation of (\ref{chi2F}), performed, this time,
with another expansion of the quantity $yx^{n}$, which underlines the role
played by some {\em fusion relations} on hypergeometric functions.

$yx^{n}$ can be written as : 
\begin{eqnarray}
\label{yxnnew}
y\,x^{n}\, &=&\,\sum_{j=0}^{\infty }{(-4w)^{j}}\cdot {n+j \choose n}
\cdot
\sin (\frac{\phi }{2})^{2j} \cdot \nonumber \\
&&{_{2}}F_{1}\Bigl(n+1/2,n+j+1;\,2n+1;\,4w\Bigr).  
\end{eqnarray}

Taking $\tilde{\chi}^{(2)}\,$ in the form (\ref{khi2expansion}), writing
$y^{2}x^{2n+2}\,$ as $yx^{n+1}\,yx^{n+1}$, using the expansion (\ref{yxnnew})
for each $yx^{n}$, the angular integration gives:
\begin{eqnarray}
\label{ok5}
\tilde{\chi}^{(2)} &=&
 4w^{4}
\sum_{n,k_1,k_2=0}^{\infty}
C(n,k_{1},k_{2})\cdot {%
_{2}}F_{1}\Bigl(n+{{3}\over{2}},n+2+k_{1};2n+3;4w\Bigr)  \nonumber \\
&&\qquad \qquad \quad \times {_{2}}F_{1}\Bigl(n+{{3}\over{2}},n+2+k_{2};2n+3;
4w\Bigr)  
\end{eqnarray}
where
\begin{eqnarray}
C(n,k_{1},k_{2}) &=& w^{2n+k_{1}+k_{2}}\cdot
(-4)^{k_{1}+k_{2}}\cdot (n+1)^{2} {\frac{ (3/2)_{k_1+k_2} }{ (3)_{k_1+k_2} }}
\nonumber \\
&& \times {n+1+k_{1} \choose k_1}{n+1+k_{2} \choose k_2}
 \nonumber 
\end{eqnarray}
and ${(N)}_{m}$ are the usual Pochhammer symbol. 
It is worth comparing (\ref{chi2F}) with (\ref{ok5}).
One sees, with this alternative expression for $\,\chi^{(2)}$
that the complexity of $\,\chi^{(2)}$ \emph{has now been encapsulated in a
single hypergeometric expression} $\,{_{2}}F_{1}\Bigl(5/2,3/2;3;16\,w^{2}
\Bigr)$. In the holonomic approach developed in this paper, the occurrence
of the prototype of holonomic functions, namely hypergeometric functions, is
quite natural. These calculations \emph{also underline the important role
the fusion identities}~\cite{gasper} (see relation (1.9)
 in~\cite{gasper}) \emph{ on hypergeometric functions}, like $\,F\,=\,\sum C\cdot
F\,F$, can play in such calculations. This is probably the sign of quite deep
structures, and symmetries, like the {\em fusion relations} encountered in CFT, or
integrable models, and deeply connected with Yang-Baxter
 relations~\cite{avan,cft}.

At this point it is worth
recalling that the fusion relation (\ref{ok5}) is totally reminiscent of,
for instance, the fusion relation (1.9) in~\cite{gasper} : 
\begin{eqnarray}
&&{}_{r+t}F_{s+u}\left[ 
\begin{array}{cc}
a_{r},\,\,  c_{t} & \\ 
\noalign{\medskip}b_{s},\,\,d_{u} & 
\end{array}
;xw\right] =\,\,\,\, \sum_{n=0}^{\infty }\frac{
(c_{t})_{n}(e_{k})_{n}(-x)^{n}}{(d_{u})_{n}(f_{m})_{n}(n+\gamma )_{n}n!} 
\nonumber  \label{reminis} \\
&&\quad \times {}_{k+t}F_{m+u+1}\left[ 
\begin{array}{cc}
n+c_{t},\,\,  n+e_{k} & \\ 
\noalign{\medskip}2n+1+\gamma ,\,\, n+d_{u},\,\,  n+\,f_{m} &
\end{array}
\,\,;\,\,x\right]  \nonumber \\
&& \quad \times {}_{m+r+2}F_{k+s}\left[ 
\begin{array}{cc}
-n,n+\gamma ,\,\, a_{r},\,\, f_{m} & \\ 
\noalign{\medskip}b_{s},\,\, e_{k} &
\end{array}
\,\,;\,\,w\right] \   \nonumber
\end{eqnarray}
where $\gamma$, $e$ and $f$ are arbitrary and $a_r$, $c_t$, $\cdots$ are
$r$, $t$, $\cdots$ upper and lower parameters.

This fusion relation, of the type $\,F(uv)\,=\,\sum C\cdot F(u)\,F(v)$,
gives a representation of a hypergeometric function of the product $\,uv$,
as a sum of products of hypergeometric functions of $\,u$ and hypergeometric
functions of $\,v$. In this respect relation (\ref{ok5}) corresponds to
particular cases of fusion relations $\,F(uv)\,=\,\sum C\cdot F(u)\,F(v)$,
with $\,u\,=\,v\,=\,4\,w$.

\section{Appendix G}
\label{appene}
To evaluate the sums in (\ref{khi3sergen}) in order to obtain our long series
for $\tilde{\chi}^{(3)}$,  substantial
computing time is spent in the computation of the coefficients of the
various functions $a(k,p)$, $g(k,p)$, $v(k,p)$, and $y_{0}x_{0}^{p}$. While
the series expansion of $a(k,p)$, and $y_{0}x_{0}^{p}$, simply read 
\begin{eqnarray}
a(k,p) &=&\sum_{j=0}^{\infty }w^{2j}\cdot {k+p+2j \choose j}
{k+p+2j \choose j+k} \\
y_{0}x_{0}^{p} &=&\sum_{j=0}^{\infty }w^{j}\cdot {2p+2j \choose j}
\end{eqnarray}%
the coefficients of the auxiliary function $g(k,p)$ are
given by recurrences of depth three with huge polynomials in $k$, $p$ and $j$.
For series generation  purposes, we have found it more efficient to use the
following inhomogeneous recurrences between the coefficients of $g(k,n)$ and 
$a(k,n)$.

Noting by $C_g(k,p,i)$ the coefficient of $w^{i}$ in the expansion of
$g(k,p)$ (and similarly for $a(k,p)$ and $v(k,p)$), the inhomogeneous recurrences read: 
\begin{eqnarray}
0 &=&\, C_g(n+2,p,i)\, -2C_g(n+1,p,i+1)+C_g(n,p,i+2)-  \nonumber \\
&&\left\{ 
\begin{array}{ll}
C_a(n+1,p,i/2+1), & \qquad i\,{\rm even} \\ 
0, & \qquad i\,{\rm odd}.
\end{array}%
\right.  \label{reccg}
\end{eqnarray}%
Once the coefficients of $g(k,p)$ are found, those of $v(k,p)$ are given by 
\begin{eqnarray}
C_v(n,p,i)\, =\, C_g(n,p,i)-C_g(n+1,p,i-1). \nonumber 
\end{eqnarray}

The following identities are useful for (\ref{reccg}): 
\begin{eqnarray}
C_g(k,p,0) &=&\, C_a(k+1,p,0),\quad \quad  C_g(k,p,1)\, =\, 2\, C_a(k+2,p,0), \\
C_g(1,p,i) &=&\, C_g(0,p,i+1)\,\, -{\frac{{1}}{{2}}}{2p+2i+4 \choose i+2}\, + \nonumber \\
&& \left\{ 
\begin{array}{ll}
{\frac{{1}}{{2}}}C_a(0,p,i/2+1), & i\,{\rm even} \\ 
&  \\ 
0, & i\,{\rm odd}
\end{array}
\right.  \label{startcg}
\end{eqnarray}%
where (\ref{startcg}) is deduced from (\ref{defv0p}).

\section{Appendix H}
Recall that the separation of $\tilde{\chi}^{(3)}$ into a relatively simple
closed form (\ref{clos}), a ``hard to compute'' series $\Xi_{h} $, and
a simpler one $\Xi_{s} $,  is made for algorithmic considerations. Although
this separation is not "natural", 
it is tempting to seek immediately, as a first step,
 a linear differential equation for (\ref{clos}) which
should be ``simpler'' 
than  $\, \tilde{\chi}^{(3)}$. Having two functions,
 $\, f$ and $\, g$, satisfying
two  homogeneous linear differential equations (like the 
second order differential equations
satisfied by $\tilde{K}$ and $\tilde{E}$ (see (\ref{KE})), 
there are some formal computer 
programs (gfun or others~\cite{gfun}) which enable one to build 
the homogeneous linear differential equation satisfied by 
the product $\, f \cdot g$ or the sum $\, f+g$, and of course 
step-by-step, a (cubic) sum of products like (\ref{clos}).
For instance, from the two order-two differential equations for
 $\tilde{K}$ and $\tilde{E}$,
one builds the order four differential equation of the product
$\tilde{K} \cdot \tilde{E}$ and the order six
differential equation of the cubic term $\tilde{K} \cdot \tilde{E}^2$.
The latter term (with a polynomial multiplying the highest derivative of degree 17),
combined with its associated algebraic expression, gives an order six
differential equation with a polynomial 
multiplying the highest derivative of degree 38.
This step by step procedure gives more and more complicated 
linear differential equations. It seems pointless to give
these extremely involved, and cumbersome, expressions. Just note that the linear 
 differential equations corresponding to the 
sum of the strictly cubic terms in $\tilde{E}$ and $\tilde{K}$ will be of order 
$\,20$, that the one for the strictly quadratic
 terms in $\tilde{E}$ and $\tilde{K}$ will be of order 
$\, 18$,  of order $\,8$ for the linear terms in $\tilde{E}$ and $\tilde{K}$, 
and, of course, of order one for the
algebraic part in  (\ref{clos}), yielding a linear differential equation
for (\ref{clos}) of order $\, 47$ and an 
{\em extremely large degree} $\, 267$ as a lower
 bound\footnote[2]{Much larger than in the seventh order Fuchsian
equation given
in~\cite{ze-bo-ha-ma-04a} for $\, \tilde{\chi}^{(3)}$ 
and recalled in the following (see (\ref{fuchs}) above).}
This means that using a bruteforce approach based on series expansions,
instead of the use of the closed form, would have required
more than $ \, 12500 \, $ coefficients for finding 
this homogeneous linear differential equation!  
As a consequence, it might look hopeless, at first sight, to seek  
a homogeneous linear differential
equation for $\tilde{\chi}^{(3)}$.

From the series point of view, let us focus on the growth of the coefficients
of the three series comprising $\tilde{\chi}^{(3)}$.
The integer coefficients of $\, w^{481}$ read, respectively, for the ``closed'', 
``simple'', ``hard'' and $\tilde{\chi}^{(3)}/8w^9$
series :
\begin{eqnarray}
&&C_{cl} \, \simeq \, 24752.108 \cdot 4^{481}, \qquad \qquad 
C_{s} \, \simeq \, -25012.475 \cdot 4^{481}, \qquad \nonumber \\
&& C_{h} \, \simeq \, 271.423 \cdot 4^{481}, \qquad \qquad \quad 
C \,  \simeq \, 11.056 \cdot 4^{481}. \nonumber
\end{eqnarray}
One thus sees that the coefficients for $\tilde{\chi}^{(3)}$ are actually 
obtained {\em from mutual cancellations of much larger integers}: the coefficients of 
$\,  \tilde{\chi}_{cl}^{(3)}$, and $\, \Xi_{s}(w)$, are
 of the same order, but of opposite signs,
their sum, $\, \, C_{cl}+C_s \simeq \,-260.366 \cdot 4^{481}$,
 is of the same order
as the coefficient
of $\, \Xi_{h}(w)$, but of opposite sign, the total sum becoming the coefficient
of $\tilde{\chi}^{(3)}$.
The coefficient of $\, w^{481}$ for the closed form  $\, \tilde{\chi}_{cl}^{(3)}$
is  $\, \simeq \, 2238.72$ times larger than the corresponding coefficient
for $\tilde{\chi}^{(3)}$!

These points show that  $\, \tilde{\chi}^{(3)}$ seen as the sum
(\ref{khi3sergen}), {\em should be much simpler
than the sum of its constituents}. This is indeed what was discovered.


\vskip 0.5cm

\begin{thebibliography}{99}


\bibitem{wu-mc-tr-ba-76}  T.T. Wu, B.M. McCoy, C.A. Tracy and E. Barouch, {\em The spin-spin
 correlation function for the two-dimensional Ising model: exact results in the scaling region}, 
1976 Phys. Rev. {\bf B 13}, 316-374 

\bibitem{nappi-78} C. R. Nappi, 1978  Nuovo Cim. A \textbf{44} 392

\bibitem{pal-tra-81} J. Palmer, C. Tracy, 1981 Adv. Appl. Math. \textbf{2} 329

\bibitem{yamada-84} K. Yamada, 1984 Prog. Theor. Phys. \textbf{71} 1416

\bibitem{yamada-85} K. Yamada, 1985 Phys. Lett. A \textbf{112} 456-458

\bibitem{nickel-99} B. Nickel, 1999 J. Phys. A: Math. Gen. \textbf{32} 3889-3906

\bibitem{nickel-00} B. Nickel, 2000 J. Phys. A: Math. Gen. \textbf{33}  1693-1711

\bibitem{ze-bo-ha-ma-04a} N. Zenine, S. Boukraa, S. Hassani, J.M. Maillard,
{\em The Fuchsian differential equation of the square lattice Ising model $%
\chi^{(3)}$ susceptibility}, 2004  J. Phys. A {\bf 37} 9651-9668 and  arXiv:math-ph/0407060

\bibitem{syo-nay-60} I. Syozi, S. Naya, 1960 Prog. Theor. Phys. \textbf{24} 829

\bibitem{HaMa88} D. Hansel and J-M. Maillard, 1988 J. Phys. A: Math. Gen. \textbf{21} 213-225

\bibitem{Dhar} D. Dhar and J-M. Maillard, 1985 J. Phys. A: Math. Gen. \textbf{18} L383-L388

\bibitem{Bax82} R.J. Baxter, 1982 J. Stat. Phys. {\bf 28}, 1-41 

\bibitem{JaMa85} M.T. Jaekel and J-M. Maillard, 1985 J. Phys. A: Math. Gen. \textbf{18} 641, 1229, 2271

\bibitem{HaMa87} D. Hansel and J-M. Maillard, {\em Formal constraints on 
series analysis on the Potts models},
 1987 Mod. Phys. Lett. {\bf B 1},  145-153

\bibitem{HaMa88Int} D. Hansel and J-M. Maillard, {\em Series analysis on the
 q-state checkerboard Potts models}, 1988 Int. Journ. Mod. Phys.  {\bf B } 1447-1462

\bibitem{yang} C. N. Yang, Phys. Rev. 1952 {\bf 85},  808-816

\bibitem{chang} C. H. Chang, Phys. Rev. 1952 {\bf 88}, 1422


\bibitem{ha-ma-oi-ve-87} D. Hansel, J.M. Maillard, J. Oitmaa, M.J. Vergakis,
J. Stat. Phys. 1987 \textbf{48}  69-80

\bibitem{gut-ent-96} A.J. Guttman, I.G. Enting, 1996 Phys. Rev. Lett. \textbf{76} 344-347


\bibitem{lipshitz-89}
L. Lipshitz, 1989 J. Algebra {\bf 122}  354

\bibitem{zeilberger-90}
D. Zeilberger, 1990 J. Comp. Appl. Math. {\bf 32} 321


\bibitem{cartier} P. Cartier, {\em Fonctions polylogarithmes, 
nombres polyz\'etas et groupes pro-unipotents},
 Ast\'erisque {\bf 282} (2002), 137-173 (S\'em. Bourbaki no. 885).

\bibitem{Tera} T. Terasoma, {\em Mixed Tate motives and multiple zeta values},
2002 Invent. Math. {\bf 149}, 339-369; preprint AG/0104231.

\bibitem{Hoff} M. E. Hoffman and Y. Ohno,
 {\em Relations of multiple zeta values and their algebraic expression},
 2003 J. Algebra {\bf 262}, 332-347; preprint QA/0010140.

\bibitem{Oi} S. Oi, {\em Representation of the Gauss hypergeometric function
 by multiple polylogarithms and relations of multiple zeta values},  preprint NT/0405162


\bibitem{ince-56} H.K. Ince, Ordinary differential equations, (Longmans,
London, 1927; also Dover Pubs., NY, 1956)

\bibitem{Schles3} U. Mugan and A. Sakka, {\em Schlesinger transformations
 for Painlev\'{e} VI equation}, 1995 J. Math. Phys. 
 {\bf 36}, 1284-1298.


\bibitem{Painl} P. Painlev\'e, 
{\em M\'emoire sur les \'equations diff\'erentielles dont l'int\'egrale
 g\'en\'erale est uniforme},
1900  Bull. Soc. Math. Phys. France {\bf 28}, 201-261.

\bibitem{Schles4} A. V. Kitaev,
{\em Special Functions of the Isomonodromy Type, Rational Transformations
of Spectral Parameter,
  and Algebraic Solutions of the Sixth Painlev\'e Equation},\\
http://arXiv.org/abs/nlin/0102020


\bibitem{Schles2} E. Barouch, B.M. McCoy and  T.T. Wu, 1973 Phys. Rev. Lett. {\bf 31}, 1409-1411 


\bibitem{chazy} J. Chazy, {\em Sur les \'equations diff\'erentielles dont l' int\'egrale 
poss\`ede une coupure essentielle mobile}, 1910 C. R. Acad. Sc. Paris {\bf 150}, 456-458  

\bibitem{chazy2} J. Chazy, {\em Sur les \'equations diff\'erentielles du
troisi\`eme et d' ordre sup\'erieur dont l' int\'egrale 
g\'en\'erale a ses points critiques fix\'es.}, 1911 Acta Math. {\bf 34}, 317-385 

\bibitem{Painl2} R. Conte, 
{\em The Painlev\'e Property, One Century Later}, CRM Series in Mathematical Physics, Springer, R. Conte
editor.

\bibitem{Kruskal} M. D. Kruskal, N. Joshi and R. Halburd,
 {\em Analytic and asymptotic methods for nonlinear
 singularity analysis: a review and extensions of tests for the Painlev\'e property}, 
 (1996) 


\bibitem{or-ni-gu-pe-01b} W.P. Orrick, B.G. Nickel, A.J. Guttmann, J.H.H.
Perk, 2001 J. Stat. Phys. \textbf{102}  795-841

\bibitem{or-ni-gu-pe-01} W.P. Orrick, B.G. Nickel, A.J. Guttmann, J.H.H.
Perk, 2001 Phys. Rev. Lett. \textbf{86} 4120-4123


\bibitem{coy-wu-80} B.M. McCoy, T.T. Wu, 1980 Phys. Rev. Lett. \textbf{45} 675-678

\bibitem{perk-80} J.H.H. Perk, 1980 Phys. Lett. A\textbf{79} 3-5

\bibitem{jim-miw-80} M. Jimbo, T. Miwa, 1980 Proc. Japan Acad. A \textbf{56}
 405; 1981 Proc. Japan Acad. A \textbf{57}  347


\bibitem{diff2} J J Rehr, G S Joyce and A J Guttmann, {\em A recurrence
 technique for confluent singularity analysis of power 
series}, 1980 J. Phys. A: Math. Gen. {\bf 13}, 1587-1602

\bibitem{diff}  M. H. Khan, {\em High-order differential approximants}, 2002 Journal of Computational 
and Applied Mathematics, {\bf 149},  Issue 2,  457-468   


\bibitem{Sasa} T. Sasaki and M. Yoshida,
	 {\em A system of differential equations in 4 variables 
	   of rank 5 invariant under the Weyl group of type $ E_6$},
	 2000 Kobe J. Math. {\bf 17}, 29-57.


\bibitem{birat} J.~M. Maillard, {\em Hyperbolic Coxeter groups, 
	symmetry group invariants for
	  lattice models in statistical mechanics and Tutte-Beraha numbers}.
	\newblock 1995  Math. Comput. Modelling {\bf 26}, 169-225 

\bibitem{birat2} S.~Boukraa, J-M. Maillard, and G.~Rollet, 
    {\em Discrete Symmetry Groups of
	  Vertex Models in Statistical Mechanics}.
	\newblock 1995 J.Stat.Phys. {\bf 78}, 1195--1251.

\bibitem{birat3} S.~Boukraa,  and  J-M. Maillard,
	 {\em Let's Baxterise}.
 	\newblock   2001 J. Stat. Phys.  {\bf 102},  641-700
	\newblock and : hep-th/0003212

\bibitem{gasper} G. Gasper, \emph{Using symbolic computer algebraic systems
to derive formulas involving orthogonal polynomials and other special
functions}, published in Orthogonal Polynomials: Theory and Practice, ed. by
P. Nevai, Kluwer Academic Publishers, Boston (1990), pp. 163-179.

\bibitem{avan} J. Avan, J-M. Maillard, M. Talon and C-M. Viallet, 
 {\em  New local relations for lattice models}.
	\newblock 1990 Int. Journ. Mod. Phys. {\bf B 4}, 1895-1912

\bibitem{cft} C. Gomez, M. Ruiz-Altaba, G. Sierra, Chapter 9 {\em Duality 
in conformal field theories} in
 {\em Quantum Groups in Two-dimensional Physics}, Cambridge
 monographs on Mathematical Physics, Cambridge university press, 1996 

\bibitem{gfun} Mgfun's project : see http://algo.inria.fr/chyzak/mgfun.html ; 
gfun - generating functions package see gfun in : http://algo.inria.fr/libraries 



\end{thebibliography}
\end{document}